\documentclass[times,twocolumn,final,longtitle]{elsarticle}

\usepackage{framed,multirow}

\usepackage{amssymb}
\usepackage{latexsym}

\usepackage{url}
\usepackage{xcolor, soul, color}
\usepackage{lipsum}  

\usepackage{hyperref,tabularx,graphicx, multirow,subcaption}
\usepackage{newtxmath, mleftright}

\definecolor{newcolor}{rgb}{.8,.349,.1}

\begin{document}


\begin{frontmatter}

\title{Placental Vessel Segmentation and Registration in Fetoscopy: Literature Review and MICCAI FetReg2021 Challenge Findings}


\author[1]{Sophia Bano\corref{cor1}\corref{cor2}}
\ead{sophia.bano@ucl.ac.uk}
\author[2,3]{Alessandro Casella\corref{cor2}} 
\author[1]{Francisco Vasconcelos} 
\author[18]{Abdul Qayyum} 
\author[18]{Abdesslam Benzinou} 
\author[5]{Moona Mazher} 
\author[6]{Fabrice Meriaudeau} 
\author[3]{Chiara Lena} 
\author[3]{Ilaria Anita Cintorrino} 
\author[3]{Gaia Romana De Paolis} 
\author[3]{Jessica Biagioli} 
\author[7]{Daria Grechishnikova} 
\author[8]{Jing Jiao} 
\author[9]{Bizhe Bai} 
\author[10]{Yanyan Qiao} 
\author[1]{Binod Bhattarai} 
\author[11]{Rebati Raman Gaire} 
\author[11]{Ronast Subedi} 
\author[12]{Eduard Vazquez} 
\author[13,19]{Szymon Płotka}  
\author[13]{Aneta Lisowska} 
\author[13,20]{Arkadiusz Sitek} 
\author[14,15]{George Attilakos} 
\author[14,15]{Ruwan Wimalasundera} 
\author[14,15,16]{Anna L. David} 
\author[17]{Dario Paladini} 
\author[15,16]{Jan Deprest} 
\author[3]{Elena De Momi} 
\author[2]{Leonardo S. Mattos} 
\author[4]{Sara Moccia} %
\author[1]{Danail Stoyanov}

\cortext[cor1]{Corresponding author.}

\cortext[cor2]{Equal contribution.}

\address[1]{Wellcome/EPSRC Centre for Interventional and Surgical Sciences (WEISS) and Department of Computer Science, University College London, UK}
\address[2]{Department of Advanced Robotics, Istituto Italiano di Tecnologia, Italy}
\address[3]{Department of Electronics, Information and Bioengineering, Politecnico di Milano, Italy}
\address[4]{The BioRobotics Institute and Department of Excellence in Robotics and AI, Scuola Superiore Sant’Anna, Italy}
\address[14]{Fetal Medicine Unit, Elizabeth Garrett Anderson Wing, University College London Hospital, UK}
\address[15]{EGA Institute for Women's Health, Faculty of Population Health Sciences, University College London, UK}
\address[16]{Department of Development and Regeneration, University Hospital Leuven,  Belgium}
\address[17]{Department of Fetal and Perinatal Medicine, Istituto "Giannina Gaslini",  Italy}
\address[18]{ENIB, UMR CNRS 6285 LabSTICC, 29238, France}
\address[5]{Department of Computer Engineering and Mathematics, University Rovira i Virgili, Spain}
\address[6]{ImViA Laboratory, University of Bourgogne Franche-Comté, France}
\address[7]{Physics Department, Lomonosov Moscow State University, Russia}
\address[8]{Fudan University, China}
\address[9]{Medical Computer Vision and Robotics Group, Department of Mathematical and Computational Sciences, University of Toronto, Canada}
\address[10]{Shanghai MicroPort MedBot (Group) Co., Ltd,}
\address[11]{NepAL Applied Mathematics and Informatics Institute for Research, Nepal}
\address[12]{Redev Technology, UK}
\address[13]{Sano Center for Computational Medicine, Poland}
\address[19]{Quantitative Healthcare Analysis Group, Informatics Institute, University of Amsterdam, Amsterdam, The Netherlands}
\address[20]{Center for Advanced Medical Computing and Simulation, Massachusetts General Hospital, Harvard Medical School, Boston, MA, United States of America}


\begin{abstract}
Fetoscopy laser photocoagulation is a widely adopted procedure for treating Twin-to-Twin Transfusion Syndrome (TTTS). The procedure involves photocoagulation pathological anastomoses to restore a physiological blood exchange among twins. The procedure is particularly challenging, from the surgeon's side, due to the limited field of view, poor manoeuvrability of the fetoscope, poor visibility due to amniotic fluid turbidity, and variability in illumination. These challenges may lead to increased surgery time and incomplete ablation of pathological anastomoses, resulting in persistent TTTS. Computer-assisted intervention (CAI) can provide TTTS surgeons with decision support and context awareness by identifying key structures in the scene and expanding the fetoscopic field of view through video mosaicking. Research in this domain has been hampered by the lack of high-quality data to design, develop and test CAI algorithms. Through the \textit{Fetoscopic Placental Vessel Segmentation and Registration (FetReg2021)} challenge, which was organized as part of the MICCAI2021 Endoscopic Vision (EndoVis) challenge, we released the first large-scale multi-center TTTS dataset for the development of generalized and robust semantic segmentation and video mosaicking algorithms with a focus on creating drift-free mosaics from long duration fetoscopy videos. For this challenge, we released a dataset of 2060 images, pixel-annotated for vessels, tool, fetus and background classes, from 18 in-vivo TTTS fetoscopy procedures and 18 short video clips of an average length of 411 frames for developing placental scene segmentation and frame registration for mosaicking techniques. Seven teams participated in this challenge and their model performance was assessed on an unseen test dataset of 658 pixel-annotated images from 6 fetoscopic procedures and 6 short clips. The challenge provided an opportunity for creating generalized solutions for fetoscopic scene understanding and mosaicking. In this paper, we present the findings of the FetReg2021 challenge alongside reporting a detailed literature review for CAI in TTTS fetoscopy. Through this challenge, its analysis and the release of multi-center fetoscopic data, we provide a benchmark for future research in this field.
\end{abstract}


\end{frontmatter}

\section{Introduction}
\label{sec:introduction}
Twin-to-Twin Transfusion Syndrome (TTTS) is a severe complication of monochorionic twin pregnancies. TTTS is characterized by an unbalanced and chronic blood transfer from one twin (the donor twin) to the other (the recipient twin) through placental anastomoses~\citep{Baschat2011}.
This shared circulation is responsible for serious complications, which may lead to profound fetal hemodynamic and cardiovascular disturbances~\citep{Lewi2013}.
In 2004, a randomized, controlled trial demonstrated that fetoscopic laser ablation of placental anastomoses in TTTS had a higher survival rate for at least one twin than other treatments, such as serial amnioreduction. Laser ablation further showed a lower incidence of complications, such as cystic periventricular leukomalacia and neurologic complications~\citep{Senat2004}. The trial included pregnancy at 16-26 weeks' gestation. Such results were confirmed for pregnancy before 17 and after 26 weeks' gestation~\citep{Baud2013}.
A description of all the steps that brought laser surgery for coagulation of placental anastomoses to be the elective treatment for TTTS can be found in~\citep{Deprest2010}.

\begin{figure*}[t!]
    
	\begin{subfigure}[b]{0.5\textwidth}
		\centering
		\includegraphics[width=0.95\textwidth]{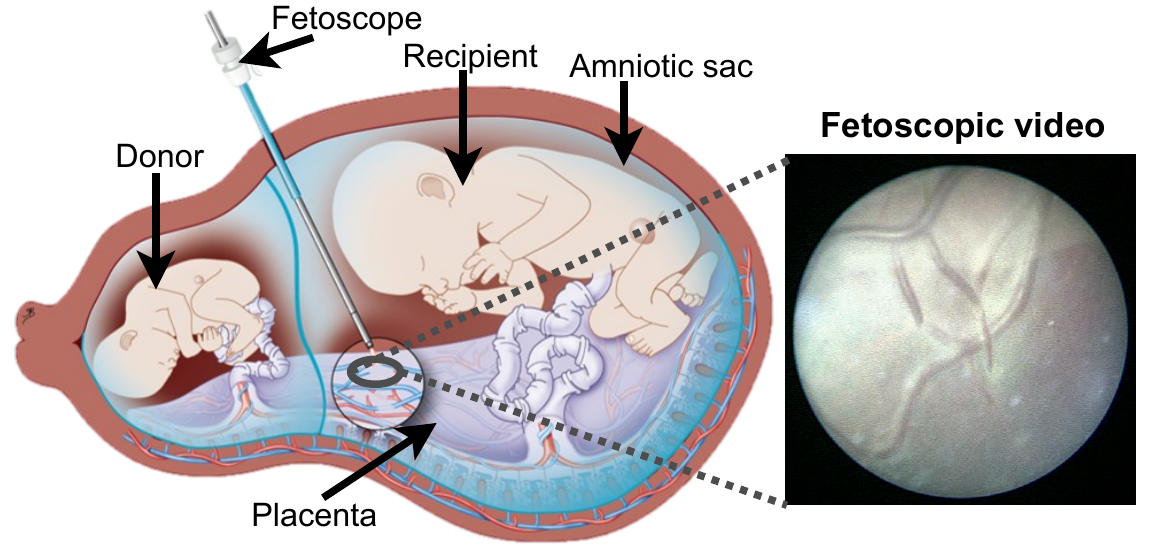}
		\caption{}	
	\end{subfigure}	
	\begin{subfigure}[b]{0.5\textwidth}
    	\centering
    	\includegraphics[width=0.8\textwidth]{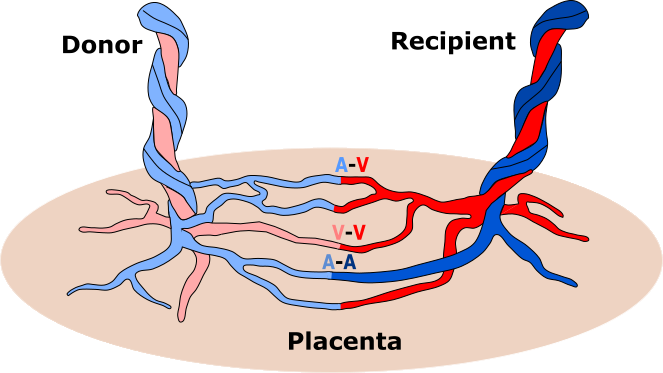}
    	\caption{}					
	\end{subfigure}	
	\caption{Illustrations of Twin-to-Twin Transfusion Syndrome. (a) shows the fetoscopic laser photocoagulation procedure, where the field of view of the fetoscope is extremely narrow. (b) shows the types of anastomoses (i) A-V: arterio-venous, (ii) V-V: veno-venous, and (iii) A-A: arterio-arterial. In the placenta, conversely from body circulatory system, arteries carries deoxygenated blood (in blue), and veins carries oxygenated blood (in red).}  
	\label{fig:TTTS_illustration}     
\end{figure*}

Fetoscopic laser photocoagulation involves the ultrasound-guided insertion of a fetoscope into the amniotic sac. Through fetoscopic camera, the surgeon identifies abnormal anastomoses and laser ablate them to regulate the blood flow between the two fetuses (as illustrated in Fig.~\ref{fig:TTTS_illustration}(a)). 
First attempts at laser coagulation included laser ablating all vessels that looked like anastomoses (a non-reproducible and operator-dependent technique), and laser ablating all vessels crossing the inter-fetus membrane (an approach that relies on the assumption that all vessels crossing the dividing membrane are pathological anastomoses)~\citep{Quintero2007}. Today, the recognized elective treatment is the \textit{selective laser photocoagulation}, which consists of the precise identification and lasering of placental pathological anastomoses. The selective treatment relies on the identification of the anastomoses (shown in Fig.~\ref{fig:TTTS_illustration}(b)) and their classification into \textit{Arterio-Venous} (from donor to recipient, AVDR, or from the recipient to donor, AVRD), \textit{Arterio-Arterial} (AA) or \textit{Veno-Venous} (VV) anastomoses. The identified AVDR anastomoses are laser ablated to regulate the blood flow between the two fetuses. 

Despite all the advancements in instrumentation and imaging for TTTS~\citep{Cincotta2016,Maselli2016}, residual anastomoses after monochorionic placentas treated with fetoscopic laser surgery still represent an issue~\citep{Lopriore2007}. This may be explained considering challenges from the surgeon's side, such as limited field of view (FoV), poor visibility and high inter-subject variability. 
In this complex scenario, computer-assisted-intervention (CAI) and surgical-data-science (SDS) methodologies may be exploited to provide surgeons with context awareness and decision support. However, the research in this field is still in its infancy, and several challenges still have to be tackled~\citep{Pratt2015}. 
These include dynamically changing views with poor texture visibility, low image resolution, non-planar view, especially in the case of the anterior placenta, occlusions due to the fetus and surgical tools, fluid turbidity and specular highlights. 

In the context of TTTS fetoscopy, approaches to anatomical landmark segmentation (inter-fetus membrane, vessel)~\citep{Casella2020,Casella2021,Sadda2019,Bano2020deep}, event detection~\citep{Vasconcelos2018,Bano2020fetnet} and mosaicking~\citep{Gaisser2018,Tella-Amo2019,Peter2018,Bano2020deep,Bano2020DSM} exist  (Sec.~\ref{sec:sota}). 
Even though fetoscopic videos have large inter- and intra-procedure variability, the majority of the segmentation and event detection approaches are validated on a small subset of in-vivo TTTS videos. 
Existing mosaicking approaches are validated only on a small subset of ex-vivo~\citep{Tella-Amo2019}, in-vivo~\citep{Peter2018,Bano2020deep} or underwater phantom sequences~\citep{Gaisser2018}. %
Intensity-based image registration~\citep{Bano2020deep,li2021globally} methods relies on placental vessel segmentation maps for registration which facilitated in overcoming some visibility challenges (e.g., floating particles, poor illumination), however, such method fails when the predicted segmentation map is inaccurate, or the vessels are inconsistent across frames or are absent from the view. Deep learning-based flow-field matching for mosaicking~\citep{tosin2022robust} has also been proposed, which results in accurate registration even in regions with poor or weak vessels but such an approach fails when the fetoscopic scene is homogenous having poor texture.

In fetoscopy, a major effort is needed to collect large, high-quality, multi-center datasets that can capture the variability of fetoscopic video. This reflects a well-known problem in the medical-image analysis community \citep{litjens2017survey} that is currently addressed by organizing international initiatives such as Grand Challenge\footnote{https://grand-challenge.org/}.

\subsection{Our Contributions}
Placental Vessel Segmentation and Registration for Mosaicking (FetReg2021)\footnote{\label{fn:fetregwebsite}FetReg challenge website: \url{https://www.synapse.org/\#!Synapse:syn25313156/}} challenge is a crowdsourcing initiative to address key problems in fetoscopy towards developing CAI techniques for providing TTTS with decision support and context awareness. 
With FetReg2021, we collected a large multi-center dataset to better capture not only inter- and intra-procedure variability but also inter-domain (data captured in two different clinical sites) variability. The FetReg2021 dataset can support developing robust and generalised models, paving the way for the translation of deep-learning methodologies in the actual surgical practice. 
The dataset is available to the research community\footnote{FetReg dataset: \url{https://www.ucl.ac.uk/interventional-surgical-sciences/weiss-open-research/weiss-open-data-server/fetreg}}, under the Creative Commons Attribution-NonCommercial-ShareAlike 4.0 International license (CC BY-NC-SA 4.0), to foster research in the field.
FetReg2021 was organized as part of the MICCAI 2021 Endoscopic Vision (EndoVis)\footnote{EndoVis Challenges: \url{https://endovis.grand-challenge.org/}} challenge, and aimed at solving two tasks: placental scene segmentation and frame registration for mosaicking. 

In this paper, we present the results and findings of the FetReg2021 challenge, in which 7 teams participated.
We further provide a detailed review of the relevant literature on CAI for fetoscopy.
To conclude, we benchmark FetReg2021 participants methods against the existing state-of-the-art in fetoscopic scene segmentation and mosaicking method.

\begin{table*}[t!]
\centering
\caption{Overview of the existing segmentation (Sec.~\ref{sec:structure_segmentation}-\ref{sec:ifm_segmentation}, fetoscopic event detection (Sec.~\ref{sec:surgical_event}) and video mosaicking methods (Sec.~\ref{sec:mosaic}). The type of dataset used in each method is also reported. Key: IFM - inter-fetus membrane; GMS - grid-based motion statistics; EMT - electromagnetic tracker.}
\label{tab:summary}
\resizebox{1.0\textwidth}{!}{
\begin{tabular}{l l l l}
\hline
\noalign{\smallskip}	
\textbf{Reference} & \textbf{Task} & \textbf{Methodology} & \textbf{Imaging type}  \\ 
\noalign{\smallskip}	
\hline 
\noalign{\smallskip}

\cite{Almoussa2011} & Vessel segmentation & Hessian filter and Neural Network trained on handcrafted features & Ex-vivo  \\

\cite{Chang2013} & Vessel segmentation & Combined Enhancement Filters & Ex-vivo (150 images) \\

\cite{Sadda2019} & Vessel segmentation & Convolutional Neural Network (U-Net) & In-vivo (345 frames from 10 TTTS procedures) \\
\cite{Bano2019} & Vessel segmentation & Convolutional Neural Network & In-vivo (483 frames from 6 TTTS procedures)  \\

\cite{Casella2020} & IFM segmentation & Adversarial Neural Network (ResNet) & In-vivo (900 frames from 6 TTTS procedures) \\

\cite{Casella2021} & IFM segmentation & Spatio-temporal Adversarial Neural Network (3D DenseNet) & In-vivo (2000 frames from 20 TTTS procedures)\footnote{Inter-Fetus Membrane Segmentation Dataset:  \url{https://zenodo.org/record/7259050}}  \\

\noalign{\smallskip}	
\hline \noalign{\smallskip}	
\cite{Reeff2006} &Mosaicking &Hybrid feature and intensity-based &In water ex-vivo placenta\\

\cite{Daga2016} &Mosaicking &Feature-based with GPU for real time computation &Ex-vivo, Phantom placenta\\

\cite{Tella2016} &Mosaicking &Combined EM and visual tracking probablistic model &
Ex-vivo w/laparoscope\& EMT\\

\cite{Gaisser2016} &Mosaicking &Deep-learned features through contrastive loss &Ex-vivo and Phantom placenta video frames\\

\cite{Yang2016} & Mosaicking &SURF features matching and RANSAC for transformation estimation &Ex-vivo and monkey placentas w/laparoscope \\

\cite{FlorisGaisser2017} &Mosaicking &Handcrafted features and LMedS for tranformation estimation &Ex-vivo, In water placenta phantom\\

\cite{Tella-Amo2018} & Mosaicking &Combined EM and visual tracking with bundle adjustment &Ex-vivo placenta w/laparoscope \& EMT \\

\cite{Gaisser2018} &Mosaicking &Extended~\cite{Gaisser2016} to detect stable vessel regions &In water placenta phantom \\ 

\cite{Sadda2018} &Mosaicking &AGAST detector with SIFT followed by GMS matching &In-vivo (\# frames/clips) \\

\cite{Peter2018} &Mosaicking &Direct pixel-wise alignment of image gradient orientations &In-vivo (\# frames/clips)\\

\cite{Tella-Amo2019} & Mosaicking &Pruning through EM and super frame generation &Ex-vivo placenta w/laparoscope \& EMT  \\

\cite{Bano2019,Bano2020deep} &Mosaicking &Deep learning-based four point registration in consecutive images &Synthetic, Ex-vivo, Phantom, In-vivo phantom)  \\

\cite{Bano2020deep}  &Mosaicking &Direct aligment of predicted vessel maps &In-vivo fetoscopy placenta data (6 procedures)~\ref{fn:fetoscopydata}\\

\cite{li2021globally} &Mosaicking &Direct aligment of predicted vessel with graph optimisation &In-vivo fetoscopy placenta data (3 procedures)~\ref{fn:fetoscopydata}\\
 
\cite{tosin2022robust}  &Mosaicking &FlowNet 2.0 with robust estimation for direct registration &Extended in-vivo fetoscopy placenta data (6 procedures)~\ref{fn:fetoscopydata}\\

\cite{casella2022learning} &Mosaicking &Learning-based keypoint matching for registration &Extended in-vivo fetoscopy placenta data (6 procedures)~\ref{fn:fetoscopydata}\\

\cite{bano2022placental} &Mosaicking &Placental vessel-guided detector-free matching for registration &Extended in-vivo fetoscopy placenta data (6 procedures)~\ref{fn:fetoscopydata}\\

\noalign{\smallskip}	
\hline \noalign{\smallskip}	
\cite{Vasconcelos2018} & Ablation detection &Binary classification using ResNet & In-vivo fetoscopy videos (5 procedures)\\

\cite{Bano2020fetnet}  & Event detection &Spatio-temporal model for multi-label classification & In-vivo fetoscopy videos (7 procedures) \\ 
\noalign{\smallskip}	
\hline
\end{tabular}
}
\end{table*}

\section{Related work}
\label{sec:sota}
This section surveys the most relevant CAI methods developed in the field of TTTS surgery. This includes anatomical structure segmentation (Sec.~\ref{sec:structure_segmentation}), mosaicking and navigation (Sec.~\ref{sec:mosaic}), and surgical event recognition (Sec.~\ref{sec:surgical_event}).   

\subsection{Anatomical structure segmentation}
\label{sec:structure_segmentation}

Image segmentation is one of the most explored task in medical image analysis.
Segmentation from intra-operative images aims at supporting surgeons' by enhancing the visibility of relevant structures (e.g., blood vessels) but presents additional challenges over anatomical image analysis due to poor texture and uncertain contours. 
Segmentation algorithms for TTTS partition mainly focuses on vessel (Sec.~\ref{sec:vessel_segmentation}) and placenta (Sec.~\ref{sec:ifm_segmentation}) segmentation, as reference anatomical structures to provide surgeons with context awareness.

\subsubsection{Placental vessel segmentation}
\label{sec:vessel_segmentation}
Since the abnormal distribution of the anastomoses on the placenta is responsible for TTTS, exploration of its vascular network is crucial during the photocoagulation procedure.
The work presented by~\cite{Almoussa2011} is among the first in the field. The work, developed and tested with ex-vivo images, combined Hessian-based filtering and a custom neural network trained on handcrafted features.
The approach was improved by~\cite{Chang2013}, which introduced a vessel enhancement filter that combined multi-scale and curvilinear filter matching. The multi-scale filter extends the Hessian filter, introducing two scaling parameters to tune vesselness sensitivity. The curvilinear filter matching refined vessel segmentation, preserving all the structures that fit in the vessel shape template defined by a curvilinear function. 
The main limitation of both methods ~\citep{Almoussa2011,Chang2013} lies in the analysis of ex-vivo images, which present different characteristics than in-vivo ones. More importantly, Hessian-based methods have been proven to perform poorly in the case of tortuous and irregular vessels~\citep{moccia2018blood}.

More recently, researchers have focused their attention on Convolutional Neural Networks (CNNs) to tackle the variability of intra-operative TTTS frames.
\cite{Sadda2019} used U-Net, achieving segmentation performance in terms of Dice Similarity Coefficient (DSC) on a dataset of 345 in-vivo fetoscopic frames of $0.55\pm0.22$.
U-Net is further explored by~\cite{Bano2020deep}, which used segmented vessels as a prior for fetoscopic mosaicking (Sec.~\ref{sec:dl_based_mos}). The authors tested several versions of U-Net, including the original version by~\cite{ronneberger2015u}, and U-Net with different backbones (i.e. VGG16, ResNet50 and ResNet101). 
The segmentation performance was evaluated on a dataset of 483 in-vivo images from six TTTS surgery, the first publicly available \footnote{\label{fn:fetoscopydata}Fetoscopy placenta dataset: \url{https://www.ucl.ac.uk/interventional-surgical-sciences/fetoscopy-placenta-data}}.

Despite the advances introduced by CNNs, the state-of-the-art methods cannot tackle the high variability of intraoperative images.
From one side, encoder-decoder architectures trained to minimize cross-entropy and DSC loss fail in segmenting poor contrasted vessels and vessels with uneven margins. Furthermore, the datasets used to train these algorithms are small and the challenges of intra-operative images, as listed in Sec.~\ref{sec:introduction}, are not always represented. 

Research in this field is strongly limited by the low availability of comprehensive expert-annotated datasets collected in different surgical settings that could encode such variability. 
This is mainly due to the low incidence of TTTS, which make systematic data collection difficult, and the lack of annotators with sufficient domain expertise to ensure clinically correct groundtruth.  

\subsubsection{Inter-fetus membrane segmentation}
\label{sec:ifm_segmentation}
At the beginning of the surgical treatment, due to the very limited FoV and poor image quality, the surgeon finds a reference for orientation within the amniotic cavity. 
The structure identified for this purpose is the inter-fetus membrane. The visibility of this membrane can be very variable, depending on the chorion characteristics, in addition to the challenges described so far in fetoscopic images.
Once located, the surgeon refers to the inter-fetus membrane as a navigation reference during placental vascular network exploration.

Automatic inter-fetus membrane segmentation has been introduced by~\cite{Casella2020} where an adversarial segmentation network based on ResNet was proposed to enforce placenta-shape constraining.
The method was tested on a dataset of 900 intraoperative frames from 6 TTTS patients with an average DSC of $91.91\%$. Despite the promising results, this method suffered when illumination was too high or low, so the membrane is barely visible in such conditions. 

The work by~\cite{Casella2020} was extended \citep{Casella2021} by exploiting dense connectivity and spatio-temporal information to improve membrane segmentation accuracy and tackle high illumination variability. 
The segmentation performance outperformed the method previously proposed when tested on the first publicly available dataset\footnote{Inter-Fetus Membrane Segmentation Dataset:  \url{https://zenodo.org/record/7259050}} of 2000 in-vivo images from 20 TTTS surgeries. 

Despite the promising results achieved in the literature, the task of inter-fetus membrane segmentation remains poorly explored, and requires further research for performance improvement and generalization. 
%
\subsection{Fetoscopic Mosaicking and Navigation}
\label{sec:mosaic}
Video mosaicking aims at generating an expanded FoV image of the scene by registering and stitching overlapping video frames. 
Video mosaicking of high-resolution images has been extensively used as navigation guidance in the context of aerial, underwater, and street view imaging and also in consumer photography to build panorama shots. However, the outputs from off-the-shelf mosaicking methods have significantly poorer quality or fail completely when applied to fetoscopy videos due to the added visibility challenges of intra-operative images. Nevertheless, fetoscopy video mosaicking remains an active research topic within the context of computer-assisted intervention. Such a technique can facilitate the surgeon during the procedure in better localization of the anastomotic sites, which can improve the procedural outcomes. 

Mosaicking for FoV expansion in fetoscopy has been explored using handcrafted feature-based and hybrid methods (Sec.~\ref{sec:feat_based_mos}), intensity-based (Sec.~\ref{sec:intensity_based_mos}), and deep learning-based (Sec.~\ref{sec:dl_based_mos}) methods. These methods are either devised for synthetic placental images, ex-vivo placental images/videos or in-vivo videos. 

\subsubsection{Handcrafted feature-based and hybrid methods}
\label{sec:feat_based_mos}
Feature-based methods involve detecting and matching features across adjacent or overlapping frames, followed by estimating the transformation between the image pairs. On the other hand, hybrid methods utilize multimodal data (combination of image and electromagnetic tracking data) or a combination of feature-based and intensity-based methods. 

Early approaches focused on accomplishing fetoscopic mosaicking from videos or overlapping a pair of images only for image registration and mosaicking.~\cite{Reeff2006} proposed a hybrid method that used classical feature detection and matching approach for first estimating the transformation of each image with respect to a reference frame, followed by global optimization by minimizing the sum of the squared differences of pixel intensities between two images. Multi-band blending was applied for seamless stitching. 
For testing the hybrid method, the authors recorded one ex-vivo placenta fixed in a hemispherical receptacle submerged in water to mimic an in-vivo imaging scenario. 
Such an experiment also allowed capturing camera calibration to remove lens distortion. A short sequence of 40 frames sampled at 3 frames per second was used for the evaluation. 
The matched feature correspondences were visually analyzed to mark them as correct or incorrect, which is a labor-intensive task. 
The generated mosaic with and without global optimization was shown for qualitative comparison. 

%
Handcrafted feature-based methods, similar to what is commonly used in high-resolution image stitching in computer vision, were also explored for fetoscopic mosaicking.
\cite{Daga2016} presented the first approach toward generating real-time mosaics. The approach considered using SIFT for feature detection and matching. For real-time computation, texture memory was used on GPU for computing extremes of the difference of Gaussian (DoG) that describes SIFT features. Planar images of ex-vivo phantom placenta recorded by mounting a fetoscope to a KUKA robotic arm were used for validating the approach. The robot was programmed to follow a spiral path that facilitated qualitative evaluation.
\cite{Yang2016} proposed a SURF feature detection and matching based approach for generating mosaics from 100 frames long sequences that captured ex-vivo phantom and monkey placentas. Additionally, pair of images correspondence failure approach was proposed based on the statistical attributes of the feature distribution and an adaptive updating mechanism for parameter tuning to recover registration failures.
\cite{FlorisGaisser2017} used different keypoint descriptors (SIFT, SURF, ORB) along with Least Median of Squares (LMedS) for estimating the transformation between overlapping pairs of images. 

Through experiments on both ex-vivo and in-water phantom sequences, the authors showed that handcrafted features returns either no features or low confidence features due to texture paucity and dynamically changing visual conditions. This leads to inaccurate or poor transformation estimation.
%

\cite{Sadda2018} proposed a feature-based method that relied on extracting AGAST corner detector~\citep{mair2010adaptive}, SIFT as descriptor and grid-based motion statistics (GMS)~\citep{bian2017gms} for refining feature matching for homography estimation. The validation was performed on 22 in-vivo fetoscopic image pairs. Additionally, in a hybrid approach by~\cite{Sadda2019}, vessel segmentation masks were also used for selecting AGAST features only around the vessel regions. However, the reported error was large mainly because of linear and single vessels in the 22 image pairs under analysis.     
Using handcrafted feature descriptors such as SIFT shows poor performance in the case of in-vivo placental videos due to the added challenges introduced by poor visibility, texture paucity and low resolution imaging.   

A few approaches used an additional electromagnetic tracker in an ex-vivo setting to design a feature-based method for improved mosaicking. 
\cite{Tella2016} and \cite{Tella-Amo2018} assumed the placenta to be planar and static and used a combination of visual and electromagnetic tracker information for generating robust and drift-free mosaics. Mosaicking performance was increased by~\cite{Tella-Amo2019}, where the pruning of overlapping frames and generation of a super frame for reducing computational time was proposed. 
An Aurora electromagnetic tracker (EMT) was mounted on the tip of a laparoscope to obtain camera pose measurements. Using this setup, a data sequence of 701 frames was captured from a phantom (i.e., a printed image of a placenta). Additionally, a synthetic sequence of 273 frames following only planar motion was also generated for quantitative evaluation. The camera pose measurements from the EMT were incorporated with frame-based visual information using a probabilistic model to obtain globally consistent sequential mosaics. It is worth mentioning that laparoscopic cameras used are considerably better than fetoscopic cameras. 
However, current clinical regulations and the limited form factor of the fetoscope hinder the use of such a tracker in intraoperative settings.

\subsubsection{Intensity-based methods}
\label{sec:intensity_based_mos}
Intensity-based image registration is an iterative process that uses raw pixel values for direct registration through first selecting features, such as edges, contours, followed by a metric, such as mutual information, cross-correlation, the sum of squared difference, absolute difference, for describing how similar two overlapping input images are and an optimizer for obtaining the best alignment through fitting a spatial transformation model.

The use of direct pixel-wise alignment of oriented image gradients for creating a mosaic was proposed by~\cite{Peter2018} that was validated on only one in-vivo fetoscopic sequence of 600 frames. An offline bag of words was used to improve the global consistency of the generated mosaic.

\cite{Bano2020deep} proposed a placental vessel-based direct registration approach. A U-Net model was trained on a dataset of 483 vessel annotated images from 6 in-vivo fetoscopy for segmenting vessels. 
The vessel maps from consecutive frames were registered, estimating the affine transformation between the frames. Testing was performed on 6 additional in-vivo fetoscopy video clips. 
The approach facilitated overcoming visibility challenges, such as floating particles and varying illumination.
However, the method failed when the predicted segmentation map is inaccurate or in views with thin or no vessels. {\cite{li2021globally} further extended this approach to propose a graph-based globally optimal image mosaicking method. The method detected loop closures with a bad-of-words scheme followed by direct image registration. Only 3 out of 6 in-invivo videos had loop closures present in them. Global refinement in alignment is then performed through G2O framwork~\citep{kummerle2011g}. }


\subsubsection{Deep learning-based methods}
\label{sec:dl_based_mos}
Existing deep learning-based methods for fetoscopic mosaicking mainly focused on training a CNN network~\citep{Bano2019, Bano2020DSM} for directly estimating homography between adjacent frames, extracting stable regions\citep{Gaisser2016} in a view, or relying on flow fields~\citep{tosin2022robust} for robust pair-wise images registration.  

A deep learning-based feature extractor was proposed by~\cite{Gaisser2016} that used similarity learning using contrastive loss when training a Siamese convolutional neural network (CNN) architecture between pairs of similar and dissimilar small patches extracted from ex-vivo placental images. The learned feature extractor was used for extracting features from pairs of overlapping images, followed by using LMedS for the transformation estimation. Due to motion blur and texture paucity that affected the feature extractor performance, the method was validated only on a short sequence (26 frames) that captured an ex-vivo phantom placenta. 
%
\cite{Gaisser2018} extended their similarity learning approach~\citep{Gaisser2016} for detecting stable regions on the vessels of the placenta. These stable regions' representation is used as features for placental image registration in an in-water phantom setting. The obtained homography estimation did not result in highly accurate registration, as the learned regions were not robust to visual variability in underwater placental scenes. 

Methods for estimating 4-point homography using direct registration with deep learning exist in computer vision literature~\citep{detone2016deep,nguyen2018unsupervised}. 
\cite{Bano2019, Bano2020DSM} extended~\citep{detone2016deep} to propose one of the first homography-based methods for fetoscopic mosaicking, which was tested on 5 diverse placental sequences, namely, synthetic sequence of 811 frames, ex-vivo placenta planar sequence of 404 frames, ex-vivo phantom placenta sequence of 681 frames, in-vivo phantom placenta sequence of 350 frames and in-vivo TTTS fetoscopic video of 150 frames. In~\citep{Bano2019,Bano2020DSM}, a VGG-like model was trained to estimate 4-point homography between two patches extracted from the same image with known transformation. Controlled data augmentation was applied to the two patches for network training. Filtering is then applied during testing to obtain the most consistent homography estimation. The proposed approach led to advancing the literature on fetoscopic mosaicking, although the network mainly focused on estimating rigid transformation (rotation and translation) between adjacent frames due to controlled data augmentation. As a result, the generated mosaics in non-planar sequences accumulated drift over time. 

More recently, deep learning-based optical flow combined with inconsistent motion filtering for robust fetoscopy mosaicking has been proposed~\citep{tosin2022robust}. Their method relied on FlowNet-v2~\citep{ilg2017flownet} for obtaining dense correspondence between adjacent frames, robust estimation using RANSAC and local refinement for removing the effect of floating particles and specularities for improved registration. Unlike~\cite{Bano2020deep} which used placental vessel prediction to drive mosaicking,~\cite{tosin2022robust} did not rely on vessels, as a result, it managed to generate robust and consistent mosaic for longer duration of fetoscopic videos. Their approach was tested on the extended fetoscopy placenta dataset from~\citep{Bano2020deep}. 

Recent computer vision literate has also introduced deep learning-based interest point descriptors \citep{detone2018superpoint,sarlin2020superglue} and detector-free dense feature matching \citep{sun2021loftr} techniques. These techniques have shown robustness in multiview feature matching. Inspired from \cite{detone2018superpoint}, \cite{casella2022learning} proposed a learning-based keypoint proposal network and an encoding strategy for filtering irrelevant keypoints based on fetoscopic image segmentation and inconsistent homographies for producing robust and drift-free fetoscopic mosaics. \cite{bano2022placental} proposed a placental vessel-guided hybrid framework for mosaicking that relies of best of \citep{Bano2020deep} and \citep{sun2021loftr}. The framework combines these two methods through a selection mechanism based on appearance consistency of placental vessels and photometric error minimization for choosing the best homography estimation between consecutive frames. \cite{casella2022learning} and \cite{bano2022placental} methods have been validated using the extended fetoscopy placenta dataset from~\citep{Bano2020deep}.

While these approach significantly improved fetoscopic mosaicking, further analysis is needed to investigate its performance in low-textured and highly non-planar placental regions.

\begin{table*}[t!]
	\centering
	\caption{Summary of the \textit{EndoVis FetReg 2021} training and testing dataset. For each video, center ID (I - UCLH, II - IGG), image resolution, the number of annotated frames (for the segmentation task), the occurrence of each class per frame and the average number of pixels per class per frame are presented. For the registration task, the number of unlabeled frames in each video clip is provided. Key: BG - background.}
	\label{tab:segreg_dataset}
	\footnotesize
	\begin{tabular}{r c >{\raggedleft\arraybackslash}m{0.6cm}
	c c >{\raggedleft\arraybackslash}m{0.7cm} >{\raggedleft\arraybackslash}m{0.7cm} >{\raggedleft\arraybackslash}m{0.7cm} >{\raggedleft\arraybackslash}m{1.0cm} >{\raggedleft\arraybackslash}m{0.8cm} >{\raggedleft\arraybackslash}m{0.7cm} >{\raggedleft\arraybackslash}m{0.7cm}c}	
	
	\hline
	\noalign{\smallskip}
	\multicolumn{12}{c}{\textbf{TRAINING DATASET}} \\
	\hline \noalign{\smallskip}	
	\textbf{Sr.} &\textbf{Video} &\textbf{Center} &\textbf{Image} &\textbf{No. of} &\multicolumn{3}{c}{\textbf{Occurrence}} &\multicolumn{4}{c}{\textbf{Occurrence}} &\textbf{Unlabel-} \\ 
	&\textbf{name} &\textbf{ID} &\textbf{Resolution} &\textbf{Labelled} &\multicolumn{3}{c}{\textbf{(frame)}}  &\multicolumn{4}{c}{\textbf{(Avg. pixels)}} &\textbf{-led clips} \\ 
	\cline{5-11} \noalign{\smallskip}	
	& & &\textbf{(pixels)} &\textbf{frames} &\textbf{Vessel} &\textbf{Tool} &\textbf{Fetus} &\multicolumn{1}{c}{\textbf{BG}} &\textbf{Vessel} &\textbf{Tool} &\textbf{Fetus} &\textbf{\# frames}\\ 
	\hline\noalign{\smallskip}	
	1. &Video001 &I &$470 \times 470$ &152 &152	&21	&11	&196463 &21493 &1462 &1482  &346 \\
	2. &Video002 &I &$540 \times 540$ &153 &153 &35 &1 &271564 &16989 &3019 &27 &259 \\ 
	3. &Video003 &I &$550 \times 550$ &117 &117 &52 &32 &260909 &27962 &3912 &9716 &541 \\ 
	4. &Video004 &II &$480 \times 480$ &100 &100 &21 &18 &212542 &14988 &1063 &1806 &388 \\ 
	5. &Video005 &II &$500 \times 500$ &100 &100 &35 &30 &203372 &34350 &2244 &10034 &722 \\ 
	6. &Video006 &II &$450 \times 450$ &100 &100 &49 &4 &171684 &28384 &1779 &653 &452 \\ 
	7. &Video007 &I &$640 \times 640$ &140 &140 &30 &3 &366177 &37703 &4669 &1052 &316 \\ 
	8. &Video008 &I &$720 \times 720$ &110 &105 &80 &34 &465524 &28049 &13098 &11729 & 295 \\ 
	9. &Video009 &I &$660 \times 660$ &105 &104 &40 &14 &353721 &68621 &7762 &5496 &265 \\ 
	10. &Video011 &II &$380 \times 380$ &100 &100	&7 &37 &128636 &8959 &184 &6621 &424 \\ 
	11. &Video013 &I &$680 \times 680$ &124 &124	&54	&21 &411713 &36907 &8085 &5695 &247 \\ 
	12. &Video014 &I &$720 \times 720$ &110 &110 &54 &14 &464115 &42714 &6223 &5348 &469 \\ 
	13. &Video016 &II &$380 \times 380$ &100 &100	&16	&20	&129888 &11331 &448 &2734 &593 \\ 
	14. &Video017 &II &$400 \times 400$ &100 &97 &20 &3 &151143 &7625	&753 &479 &490 \\ 
	15. &Video018 &I &$400 \times 400$ &100 &100	&26	&11	&139530 &15935 &1503 &3032 &352 \\ 
	16. &Video019 &II &$720 \times 720$ &149 &149	&15	&31	&470209 &38513 &1676 &8002 &265 \\ 
	17. &Video022 &II &$400 \times 400$ &100 &100	&12	&1	&138097 &21000 &650 &253 &348 \\ 
	18. &Video023 &II &$320 \times 320$ &100 &92 &14 &8 &94942 &6256 &375 &828 &639 \\ 
	\hline\noalign{\smallskip}
	\multicolumn{3}{c}{\textbf{All training videos}} &\textbf{2060} &2043 &581 &293 &4630229 &467779 &58905 &74987 &\textbf{7411} \\ 
	\hline\noalign{\smallskip}\noalign{\smallskip}
	\multicolumn{12}{c}{\textbf{TESTING DATASET}} \\
	\hline \noalign{\smallskip}	
	19. &Video010 &II &$622 \times 622$ &100 &92	&7	&28	&341927	&40554	&1726	&19410  &320 \\ 
	20. &Video012 &II &$320 \times 320$ &100 &100	&54	&0 &95845	&5132	&1422	&0 &507 \\ 
	21. &Video015 &I &$720 \times 720$ &125 &124	&83	&28 &452552	&47221	&12082	&6545 &530 \\ 
	22. &Video020 &I &$720 \times 720$ &123 &100	&15	&1 &436842	&59884	&15259	&6415 &307 \\ 
	23. &Video024 &II &$320 \times 320$ &100 &110	&72	&13 &203372 &34350 &2244 &10034 &269 \\ 
	24. &Video025 &I &$720 \times 720$ &110 &648	&320	&83 &459947	&43189	&9801	&5464 &272 \\ 
	\hline\noalign{\smallskip}
	\multicolumn{3}{c}{\textbf{All testing videos}} &\textbf{658} &2043 &581 &293 &1880090 &205009	&40638	&37879 &\textbf{2205} \\ 
	\hline\noalign{\smallskip}

	\end{tabular}
\end{table*}

\subsection{Surgical event recognition} 
\label{sec:surgical_event}

TTTS laser therapy has a relatively simple workflow with an initial inspection of the vasculature and placenta surface to identify and visualize photocoagulation targets. 
Fetoscopic laser therapy is conducted by photocoagulation of each identified target in sequence. 
Automatic identification of these surgical phases and surgical events is an essential step towards general scene understanding and tracking of the photocoagulation targets. This identification can provide temporal context for tasks such as segmentation and mosaicking. It could also provide prior to finding the most reliable images for registration (before ablation) or identify changes in the appearance of the scene (after ablation). 

The CAI literature has hardly explored event detection or workflow analysis methods.
\cite{Vasconcelos2018} used a ResNet encoder to detect ablation in TTTS procedures, additionally also indicating when the surgeon is ready for ablating the target vessel. The method was validated on 5 in-vivo fetoscopic videos. 
\cite{Bano2020fetnet} combined CNNs and recurrent networks for the spatio-temporal identification of fetoscopic events, including clear view, occlusion (i.e., fetus or working channel port in the FoV), laser tool presence, and ablating laser tool present. The method was effective in identifying clear view segments~\citep{Bano2020fetnet} suitable for mosaicking and was validated on 7 in-vivo fetoscopic videos. Due to inter- and intra-case variability present in fetosopic videos, evaluation on a larger dataset is needed to validate the generalization capabilities of the current surgical event recognition methods.

\section{The FetReg Challenge: Dataset, Submission, Evaluation}

In this section, we present the dataset of the \textit{EndoVis FetReg 2021} challenge and its tasks (Sec.~\ref{sec:dataset}), the evaluation protocol designed to assess the performance of the participating methods (Sec.~\ref{sec:protocol}) and an overview of the challenge setup and submission protocol(Sec.~\ref{sec:overview}).   

\subsection{Dataset and Challenge Tasks}
\label{sec:dataset}

The \textit{EndoVis FetReg 2021} challenge aims at advancing the current state-of-the-art in placental vessel segmentation and mosaicking~\citep{Bano2020deep} by providing a benchmark multi-center large-scale dataset that captured variability across different patients and different clinical institutions. We also aimed to perform out-of-sample testing to validate the generalization capabilities of trained models. The participants were required to complete two sub-tasks which are critical in fetoscopy, namely: 

\begin{itemize}
    \item \textbf{Task 1: Placental semantic segmentation}: The participants were required to segment four classes, namely, background, vessels, tool (ablation instrument, i.e. the tip of the laser probe) and fetus, on the provided dataset. Fetoscopic frames from 24 TTTS procedures collected in two different centers were annotated for the four classes that commonly occur during the procedure. This task was evaluated on unseen test data (6 videos) independent of the training data (18 videos). The segmentation task aimed to assess the generalization capability of segmentation models on unseen fetoscopic video frames.
    
    \item \textbf{Task 2: Registration for Mosaicking}: The participants were required to perform the registration of consecutive frames to create an expanded FoV image of the fetoscopic environment. Fetoscopic video clips from 18 multi-center fetoscopic procedures were provided as the training data. No registration annotations were provided as it is not possible to get the groundtruth registration during the in-vivo clinical fetoscopy. The task was evaluated on 6 unseen video clips extracted from fetoscopic procedure videos, which were not part of the training data. The registration task aimed to assess the robustness and performance of registration methods for creating a drift-free mosaic from unseen data. 
\end{itemize}


The \textit{EndoVis FetReg 2021} dataset is unique as it is the first large-scale fetoscopic video dataset of 24 different TTTS fetoscopic procedures. The videos contained in this dataset are collected from two fetal surgery centers across Europe, namely,
\begin{itemize}
    \item Center I: Fetal Medicine Unit, University College London Hospital (UCLH), London, UK,
    \item Center II: Department of Fetal and Perinatal Medicine, Istituto "Giannina Gaslini" (IGG), Genoa, Italy,
\end{itemize}
Both centers contributed with 12 TTTS fetoscopic laser photocoagulation videos each. A total of 9 videos from each center (18 videos in total) form the training set, while 3 videos from each center (6 videos in total) form the test set. Alongside capturing the intra-case and inter-case variability, the multi-center data collection allowed capturing the variability that arises due to different clinical settings and imaging equipment at different clinical sites. 
At UCLH, the data collection was carried out as part of the GIFT-Surg\footnote{GIFT-Surg project: \url{https://www.gift-surg.ac.uk/}} project. The requirement for formal ethical approval was waived, as the data were fully anonymized in the corresponding clinical centers before being transferred to the organizers of the \textit{EndoVis FetReg 2021} challenge. \looseness=-1

\begin{figure*}[ht]
\centering
\includegraphics[width=1.0\linewidth]{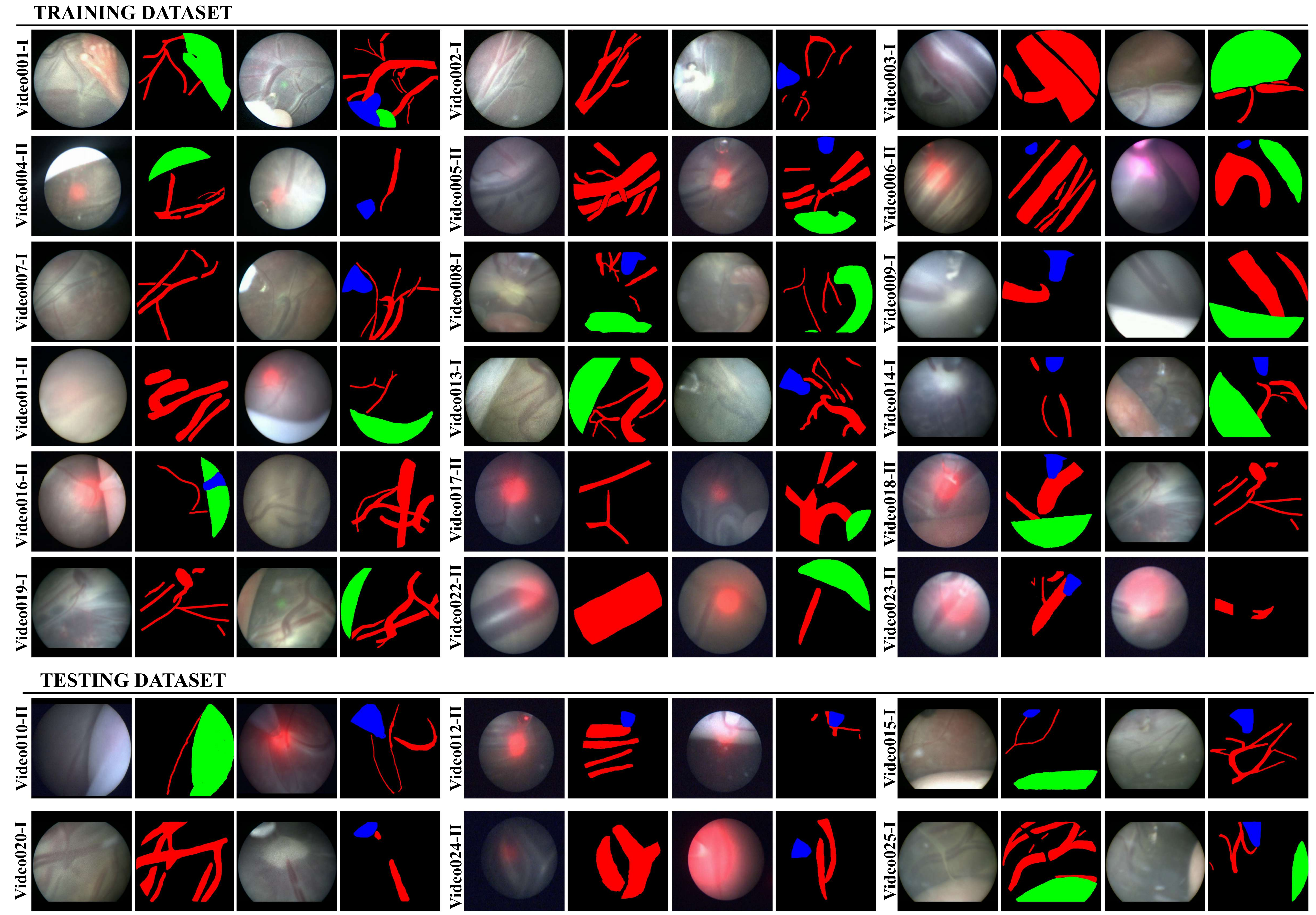}
\caption{Representative images from training and test datasets along with the segmentation annotations (groundtruth). Each center ID is also indicated next to video name (I - UCLH, II - IGG) for visual comparison of variabilities between the two centers.}
\label{fig:seg_rep}
\end{figure*}

\begin{figure*}[h]
\centering
\includegraphics[width=1.0\linewidth]{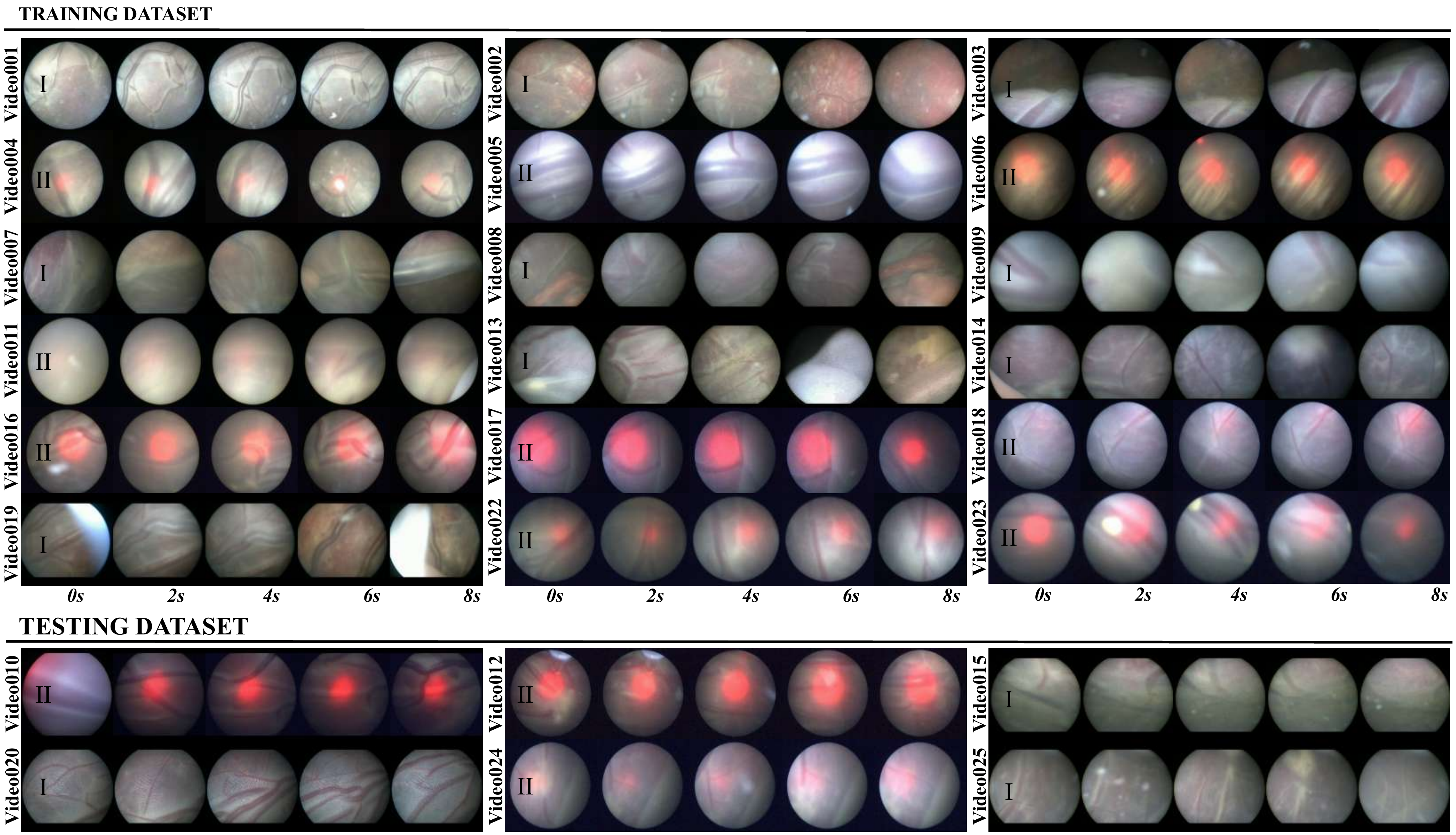}
\caption{Representative frames from training and test datasets at every 2~\textit{seconds}. These clips are unannotated and the length of each clip mentioned in Table~\ref{tab:segreg_dataset}. Center ID is also marked on each video sequence (I - UCLH, II - IGG) for visual comparison of the data from the two different centers.}  
\label{fig:seq_rep}
\end{figure*}

Table~\ref{tab:segreg_dataset} summarizes \textit{EndoVis FetReg 2021} dataset characteristics and also indicates the center from which it is acquired. 
Videos from the two centers varied in terms of the resolution, imaging device and light source. The videos from UCLH are of higher resolution (minimum resolution: $470 \times 470$, maximum resolution: $720 \times 720$) with majority videos having 720p resolution compared to IGG (minimum resolution: $320 \times 320$, maximum resolution: $622 \times 622$) videos with majority having 400p or lower resolution. From Fig.~\ref{fig:seg_rep} and Fig.~\ref{fig:seq_rep}, we can observe that most of the IGG center videos have a dominant red spotlight light visible with most views appearing to be very close to the placental surface. On the other hand, no domain light reflection is visible in any of the UCLH center videos and the imaging device captured relatively wider view compared to the IGG videos. Additionally, the frame appearance and quality changes in each video due to the large variation in intra-operative environment among different cases. Amniotic fluid turbidity resulting in poor visibility, artefacts introduced due to spotlight light source, low resolution, texture paucity, non-planar views due to anterior placenta imaging, are some of the major factors that contribute to increase the variability in the data from both centers. Large intra-case variations can also be observed from Fig.~\ref{fig:seg_rep} and Fig.~\ref{fig:seq_rep}. All these factors contribute toward limiting the performance of the existing placental image segmentation and registration methods~\citep{Bano2020deep,Bano2019,Bano2020DSM}. 
The \textit{EndoVis FetReg 2021} challenge provided an opportunity to make advancements in the current literature by designing and contributing novel segmentation and registration methods that are robust even in the presence of the above-mentioned challenges.
Further details about the segmentation and registration datasets are provided in following sections.

\begin{figure}[t!]
	\begin{subfigure}[b]{0.5\textwidth}
		\centering
		\includegraphics[ width=\textwidth]{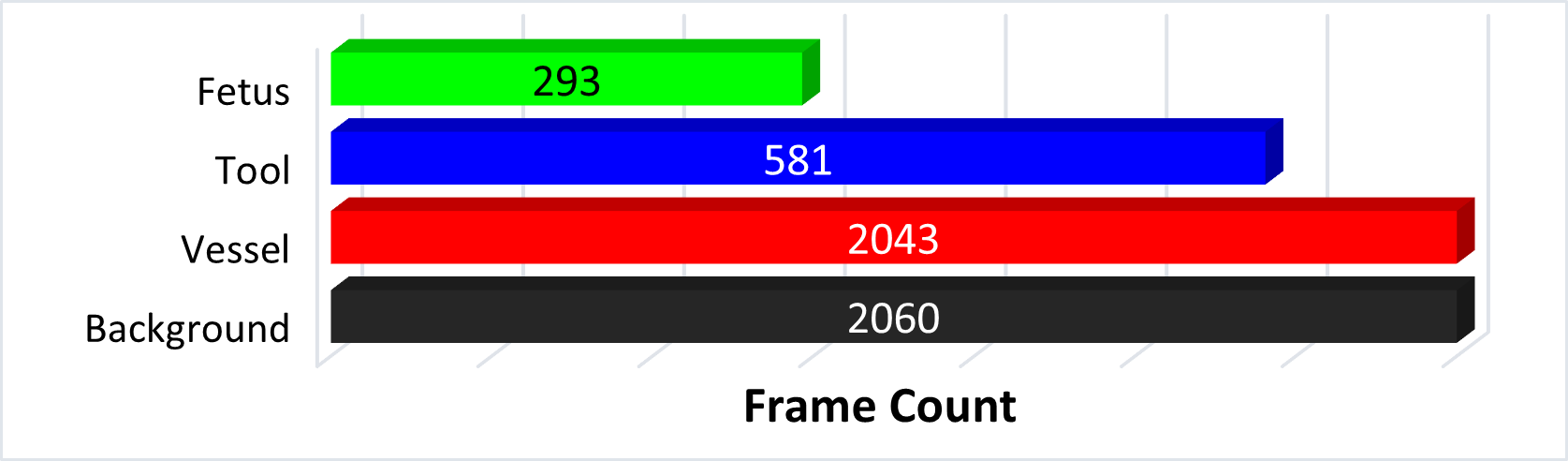}
		\caption{Occurrence per frame}	
	\end{subfigure}	
	\begin{subfigure}[b]{0.5\textwidth}
    	\centering
    	\includegraphics[width=\textwidth]{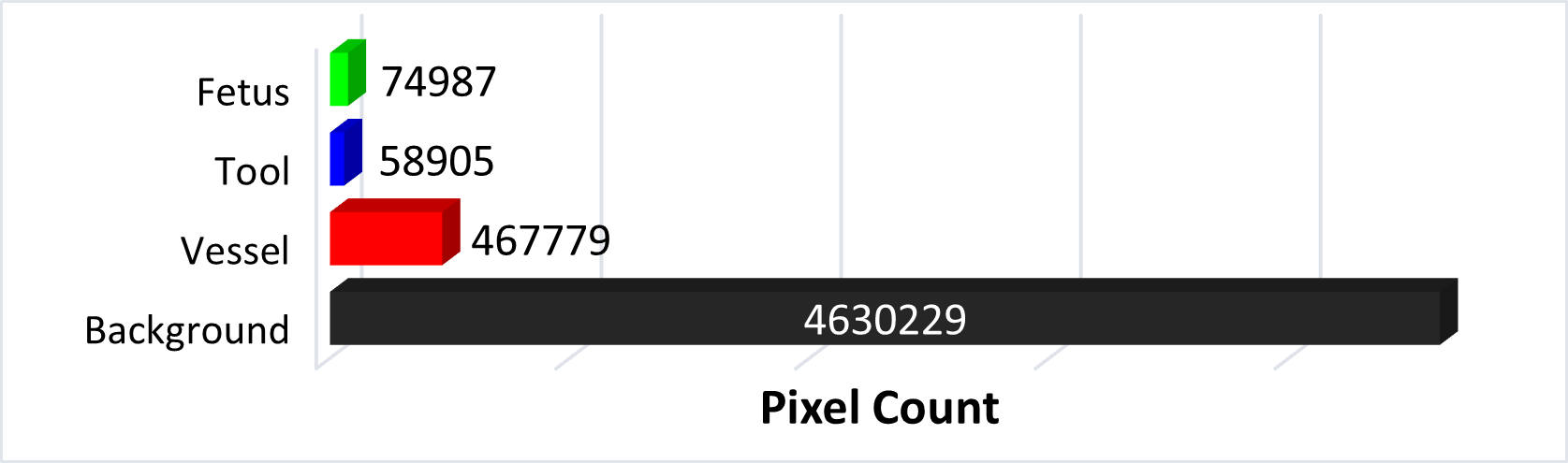}
    	\caption{Occurrence average pixel}					
	\end{subfigure}	
	\caption{Training dataset distribution: (a) and (b) segmentation classes and their overall distribution in the segmentation data.}
	\label{fig:train_data}     
\end{figure}

\begin{figure}[t!]
	\begin{subfigure}[b]{0.5\textwidth}
    	\centering
    	\includegraphics[width=\textwidth]{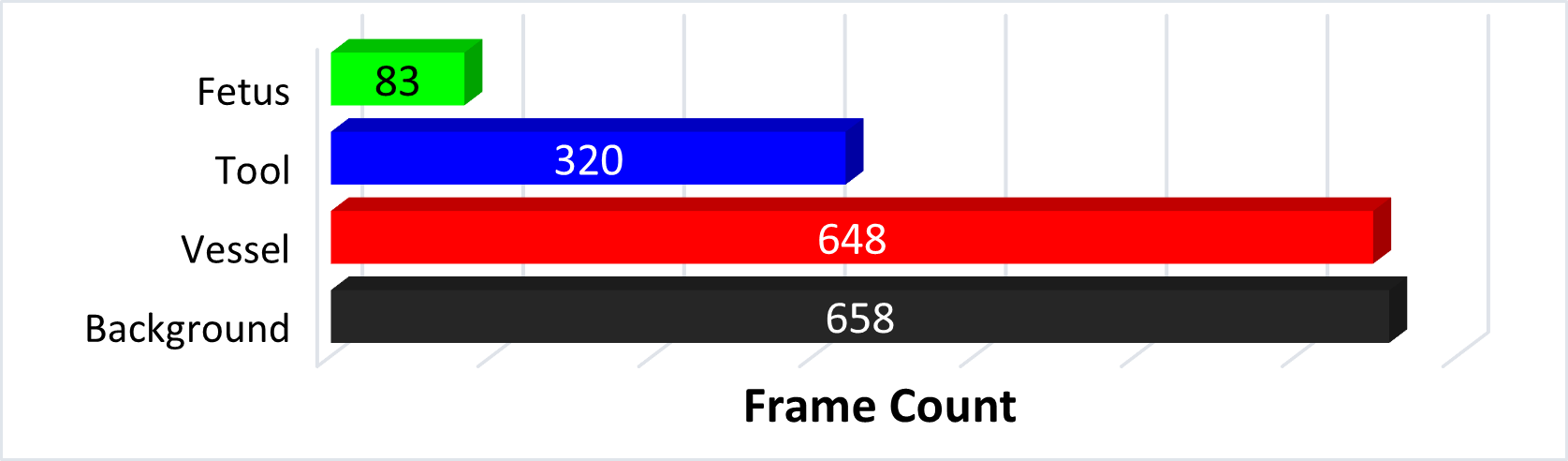}
    	\caption{Occurrence per frame}					
	\end{subfigure}	
	\begin{subfigure}[b]{0.5\textwidth}
    	\centering
    	\includegraphics[width=\textwidth]{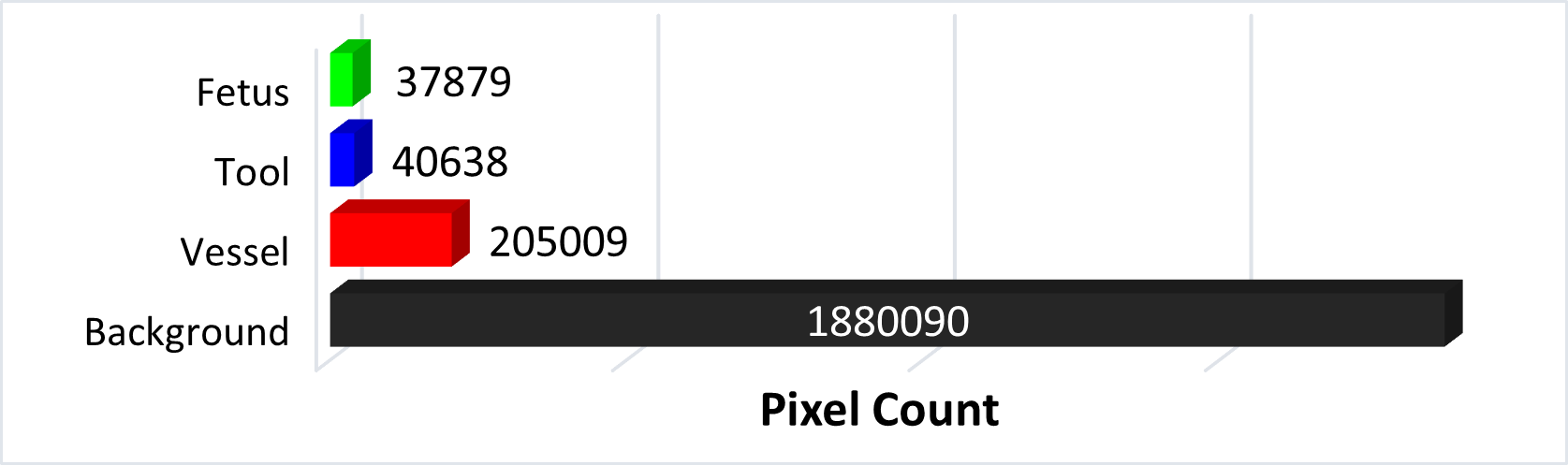}
    	\caption{Occurrence average pixel}					
	\end{subfigure}	
	\caption{Testing dataset distribution: (a) and (b) segmentation classes and their overall distribution n the segmentation data.} 
	\label{fig:test_data}     
\end{figure}

\subsubsection{Dataset for placental semantic segmentation}

Fetoscopy videos acquired from the two different fetal medicine centers were first decomposed into frames, and excess black background was cropped to obtain squared images capturing mainly the fetoscope FoV. From each video, a subset of non-overlapping informative frames (in the range 100-150) is selected and manually annotated. All pixels in each image are labelled with background (class 0), placental vessel (1), ablation tool (2) or fetus class (3). Labels are mutually exclusive.

Annotation of 7 out of 24 videos was performed by four academic researchers and staff members with a solid background in fetoscopic imaging. Additionally, annotation services are obtained from Humans in the Loop (HITL)\footnote{Humans in the Loop: \url{https://humansintheloop.org/}} for a subset of videos (17 out of 24 videos), who provided annotators with clinical background. Each image was annotated once following a defined annotation protocol. All annotations were then verified by two academic researchers for their correctness and consistency. Finally, two fetal medicine specialists verified all the annotations to confirm the correctness and consistency of the labels. The publicly available Supervisely\footnote{Supervisely: a web-based annotation tool: \url{https://supervise.ly/}} platform was used for annotating the dataset.

The \textit{FetReg} train and test dataset for the segmentation task contains 2060 and 658 annotated images from 18 and 6 different in-vivo TTTS fetoscopic procedures, respectively. 
Figure~\ref{fig:train_data}(a) and Fig.~\ref{fig:train_data}(b) show the overall class occurrence per frame and class occurrence in average pixels per frame on the training dataset. The same for test dataset is shown in Figure.~\ref{fig:test_data}(a) and Fig.~\ref{fig:test_data}(b). 
Note that the frames present different resolutions as the fetoscopic videos are captured at different centers with different facilities (e.g., device, light scope). 
The dataset is highly unbalanced: \textit{Vessel} is the most frequent class while \textit{Tool} and \textit{Fetus} are presented only in a small subset of images corresponding to 28\% and 14\%, respectively of the training dataset and 48\% and 13\% of the test dataset. When observing the class occurrence in average pixels per image, the \textit{Background} class is the most dominant, with \textit{Vessel}, \textit{Tool} and \textit{Fetus} occur 10\%, 0.13\% and 0.16\% in train dataset and 11\%, 0.22\%, and 0.20\% in test dataset, respectively.    

Figure~\ref{fig:seg_rep} shows some representative annotated frames from each video. Note that the frame appearance and quality change in each video due to the large variation in the intra-operative environment among different cases. Amniotic fluid turbidity resulting in poor visibility, artifacts introduced due to spotlight light source and reddish reflection introduced by the laser tool, low resolution, texture paucity, and non-planar views due to anterior placenta imaging are some of the major factors that contribute to increase the variability in the data. Large intra-case variations can also be observed from these representative images. All these factors contribute toward limiting the performance of the existing placental image segmentation and registration methods~\citep{Bano2020deep,Bano2019,Bano2020DSM}. 
The \textit{EndoVis FetReg 2021} challenge provided an opportunity to make advancements in the current literature by designing and contributing novel segmentation and registration methods that are robust even in the presence of the above-mentioned challenges.

\subsubsection{Dataset for registration for mosaicking}
A typical TTTS fetoscopy surgery takes approximately 30 minutes. Only a sub-set of fetoscopic frames is suitable for frame registration and mosaicking because fetuses, laser ablation fibre, and working channel port can occlude the field-of-view of the fetoscope. Mosaicking is mainly required in occlusion-free video segments that capture the surface of the placenta~\citep{Bano2020fetnet} as these are the segments in which the surgeon is exploring the intraoperative environment to identify abnormal vascular connections. Expanding the FoV through mosaicking in these video segments can facilitate the procedure by providing better visualization of the environment. 

For the registration for the mosaicking task, we have provided one video clip per video for all 18 procedures in the training dataset. Likewise, one clip per video from all 6 procedures in the test dataset is selected for testing and validation.
These frames are neither annotated with segmentation labels nor have registration groundtruth. The number of frames in each video clip is reported in Table~\ref{tab:segreg_dataset} for training and test dataset. Representative frames from each clip are shown in \ref{fig:seq_rep}.

Representative frames every 2 seconds from some video clips are shown in Fig.~\ref{fig:seq_rep}. Observe the variability in the appearance, lighting conditions and image quality in all video clips. Even though there is no noticeable deformation in fetoscopic videos, which is usually thought to occur due to breathing motion, the views can be non-planar as the placenta can be anterior or posterior. 
Moreover, there is no groundtruth camera motion and scene geometry that can be used to evaluate video registration approaches for in-vivo fetoscopy. In Section~\ref{sec:reg_metrics}, we detail how this challenge is addressed with an evaluation metric that is correlated with good quality, consistent, and complete mosaics~\citep{Bano2020deep}.  

\subsection{Evaluation protocol}
\label{sec:protocol}

\subsubsection{Segmentation Evaluation}

Intersection over union ($\textrm{IoU}$) is another most commonly used metric for evaluating segmentation algorithms which measure the spatial overlap between the predicted and groundtruth segmentation masks as: 
\begin{equation}
\label{eq: iou}
    \textrm{IoU} =\frac{\textrm{TP}}{\textrm{TP}+\textrm{FP}+\textrm{FN}}
\end{equation}
where $\textrm{TP}$ are the correctly classified pixels belonging to a class, $\textrm{FP}$ are the pixels incorrectly predicted in a specific class, and $\textrm{FN}$ are the pixels in a class incorrectly classified as not belonging to it. 
For evaluating the performance of segmentation models (Task 1), we compute for each frame provided in the test set the mean Intersection over Union ($\textrm{mIoU}$) per class between the prediction and the manually annotated segmentation masks. 
Overall mean $\textrm{mIoU}$ over all classes and all test samples is also computed and used for ranking different methods under comparison. 

\begin{figure}[t!]
\centering
\includegraphics[width=1.0\columnwidth]{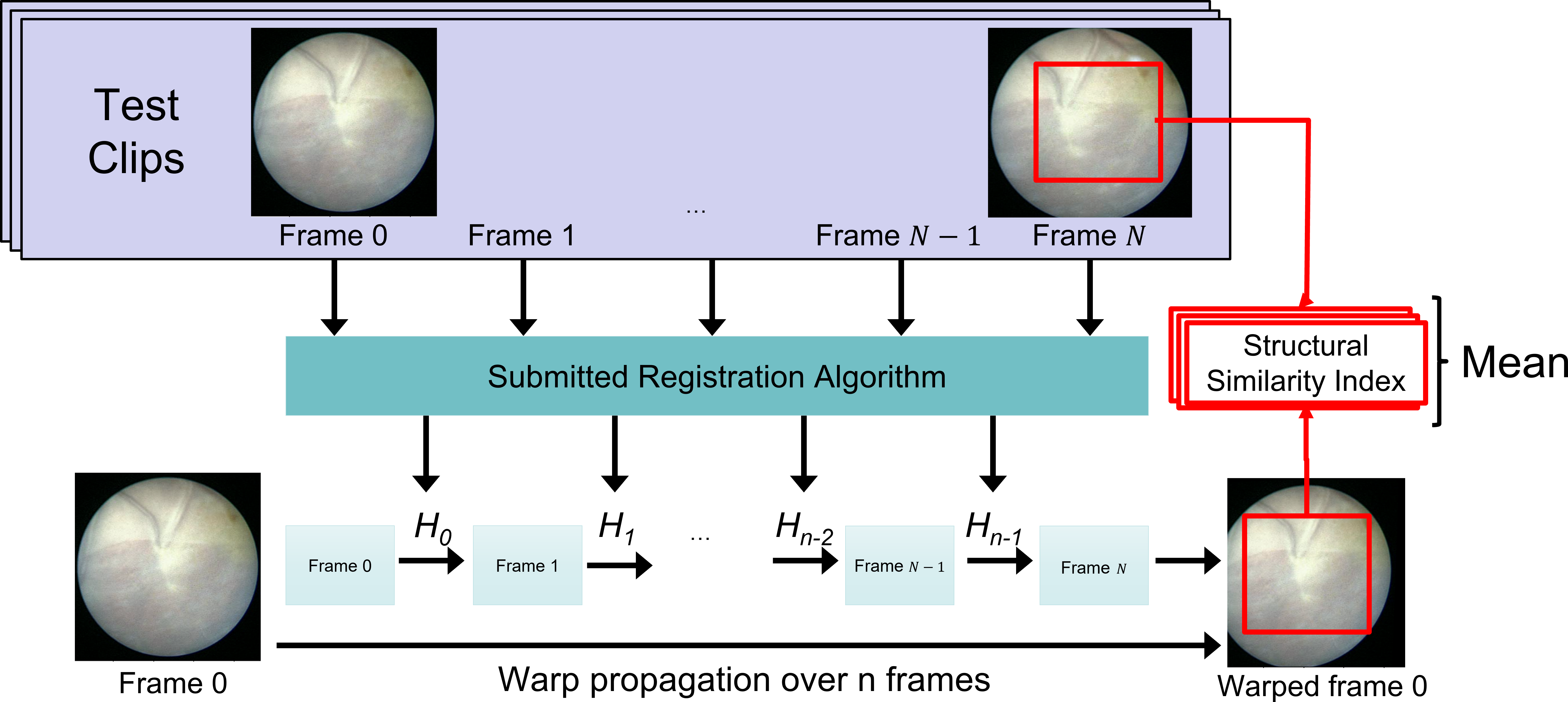}
\caption{Illustration of the N-frame SSIM evaluation metric from~\cite{Bano2020deep} } 
\label{fig:reg_eval_met}
\end{figure}

\subsubsection{Frame Registration and Mosaicking Evaluation}
\label{sec:reg_metrics}
For evaluating homographies and mosaics (Task 2), we use the evaluation metric presented by~\cite{Bano2020deep} in the absence of groundtruth. 
The metric that we referred as $N$-frame structural similarity index measure (SSIM) aims to evaluate the consistency in the adjacent frames. A visual illustration of the $N$-frame SSIM metric is presented in Fig.~\ref{fig:reg_eval_met}.
Given $N$ consecutive frames and a set of $N-1$ homographies $\{H_1, H_2,..., H_{N-1}\}$, we evaluate the consistency between them. 
The ultimate clinical goal of fetoscopic registration is to generate consistent, comprehensible and complete mosaics that map the placental surface and guide the surgeon. Considering adjacent frames will have a large overlap along them, we evaluate the registration consistency between pairs of non-consecutive frames $N$ frames apart that have a large overlap in the FoV and present a clear view of the placental surface. 
Consider a source image $I_{i}$, a target image $I_{i+n}$, and a homography transformation ${H}_{i \rightarrow i+n}$ between them, we define the consistency $s$ between these two images as:
\begin{equation}
    s_{i \rightarrow i+n} =\mathrm{sim}(w(\tilde{I}_{i},H_{i \rightarrow i+n}),\tilde{I}_{i+n})
\end{equation}
where $\mathrm{sim}$ is an image similarity metric that is computed based on the target image and warped source image, and $\tilde{I}$ is a smoothed version of the image $I$. \textit{Smoothing} $\tilde{I}$ is obtained by applying a $9\times9$ Gaussian filter with a standard deviation of 2 to the original image $I$. 
This is fundamental to make the similarity metric robust to small outlier (e.g., particles) and image discretization artifacts. For computing the \textit{similarity}, we start by determining the overlap region between the target $\tilde{I}$ and the warped source $w(\tilde{I}_{i},{H}_{i \rightarrow i+n})$, taking into account their circular edges. 
If the overlap contains less than $25\%$ of $\tilde{I}$, we consider that the registration failed, as there will be no such cases in the evaluation pool. A rectangular crop fits the overlap, and the SSIM is calculated between the image pairs after being smoothed, warped, and cropped. 

\begin{figure*}[t]
\centering
\includegraphics[width=1.0\textwidth]{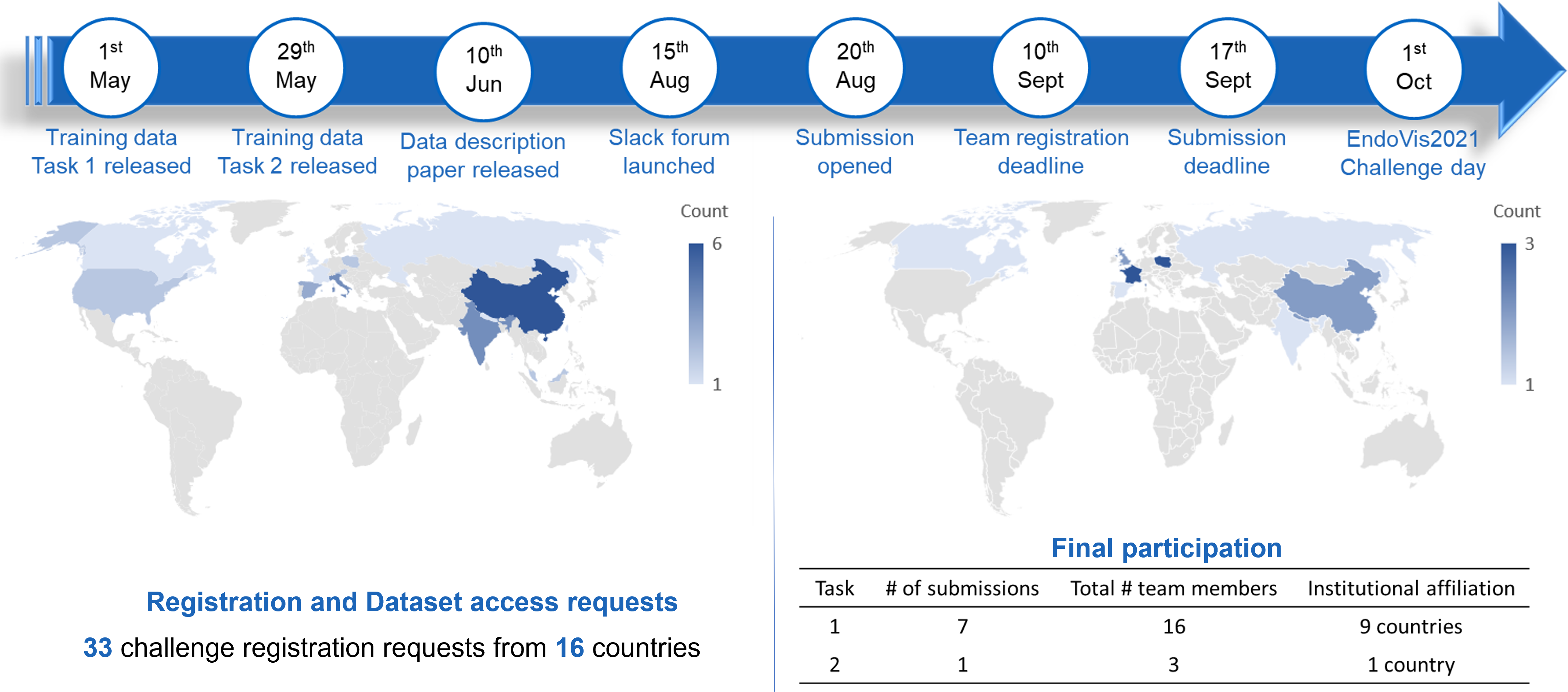}
\caption{FetReg2021 timeline and challenge participation statistics. } 
\label{fig:challenge_participation}
\end{figure*}

\subsection{Challenge Organization and Timeline}
\label{sec:overview}
The \textit{FetReg 2021} challenge is a crowdsourcing initiative that was organized by Sophia Bano (University College London, London, UK), Alessandro Casella (Istituto Italiano di Tecnologia and Politecnico di Milano, Italy), Francisco Vasconcelos (University College London, London, UK), Sara Moccia (Scuola Superiore Sant’Anna, Italy) and Danail Stoyanov (University College London, London, UK). The \textit{FetReg 2021} challenge was organized as part of the EndoVis challenge series, which is led by Stefanie Speidel (German Cancer Research Center, Heidelberg, Germany), Lena Maier-Hein (German Cancer Research Center, Heidelberg, Germany) and Danail Stoyanov (University College London, London, UK).

The FetReg challenge was organized according to The Biomedical Image Analysis Challenges (BIAS)~\citep{MAIERHEIN2020101796} reporting guideline to enhance the quality and transparency of health research.

The challenge timeline and submission statistics are presented in Fig.~\ref{fig:challenge_participation}. The challenge was announced on April 1st 2021, through the FetReg2021 Synapse~\ref{fn:fetregwebsite} website. The training dataset for task 1 and task 2 was released on May 1st and 29th, respectively. No restrictions were imposed on using additional publicly available datasets for training.
A challenge description paper~\citep{bano2021fetreg} that also included baseline method evaluation was also published on June 10th. All the details regarding the baseline methods (i.e., architecture, algorithms, and training settings) for segmentation and registration have been publicly disclosed along with its release.
Additionally, a slack support forum was launched for faster communication with the participants. 

Docker submission was opened on August 20th 2021, followed by the team registration deadline of September 10th, and the final submission deadline was set to September 17th. Members of the organizers' department may participate in the challenge, but were not eligible for awards.

Through the FegReg website, it was announced since the start of the challenge that the top three performing methods will be announced publicly during the challenge day, and the top method for each task will be awarded with a prize from the sponsors.
The remaining teams could decide whether their identity should be publicly revealed or not (e.g., in the challenge publication). All participating team, whose method achieved an overall mIoU of over 0.25 were included in this joint publication. Only one team was excluded as their method resulted in an extremely low mIoU of 0.060 on the test set (see Section \ref{sec:team_summary}.) 

\begin{figure}[t]
\centering
\includegraphics[width=0.7\columnwidth]{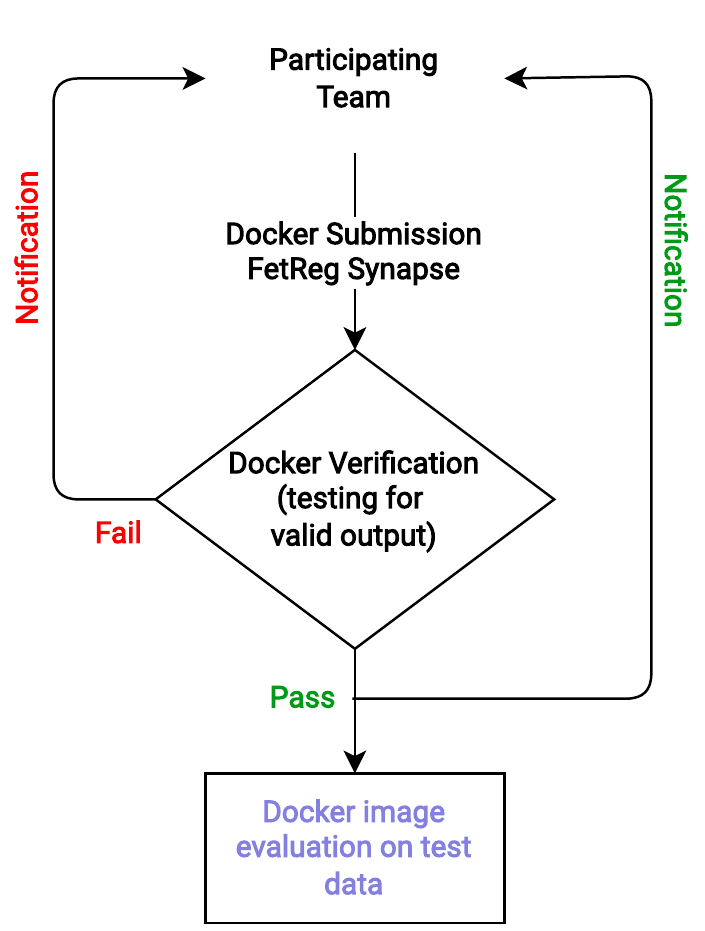}
\caption{FetReg2021 submission protocol illustrating the docker image verification protocol. } 
\label{fig:sub_proto}
\end{figure}

The test dataset was not made available to the challenge participants to keep the comparison fair and avoid misuse of the test data during training. Each participating team was required to make submissions as a docker container that accepts a path to a folder containing video frames from a patient as input and outputs a segmentation mask as an image (for task 1) or a text file with relative homography matrix (for task 2). Only fully automatic algorithms are allowed to participate in the challenge. 

The teams could submit multiple docker dockers during the submission time (from August 20th to September 17th 2021) to check the validity of the docker. We provided the participants with docker examples for both tasks along with detailed submission guidelines through FetReg2021 GitHub repository\footnote{FetReg2021 GitHub: \url{https://github.com/sophiabano/EndoVis-FetReg2021}}. 
The docker submission protocol is illustrated in Fig.~\ref{fig:sub_proto}. Each participating team submitted their docker through the Synapse platform. The submitted docker was verified for the validity of their output structure, i.e., they follow the same output format as requested and needed for the evaluation. Each participating team was then informed whether their submission passed the validity test. Each team was allowed to submit multiple dockers. However, only the last valid docker submission was used in the final evaluation.

We received 33 challenge registration requests from 16 different countries. A total of 13 team registration requests with a total number of 22 team members were received. For task 1, final submissions were received from 7 teams having 16 participants. For task 2, one submission was received, probably because of the challenging nature of this task. 

\section{Summary of methods proposed by participating teams}
\label{sec:team_summary}
In total, 7 teams participated in the challenge. Out of these, one team did not qualify to be included in this article as the achieved performance was extremely low with a \textit{mIoU} of 0.060. In this section, we summarize the methodology proposed by each participating team.

\begin{figure*}[ht]
\centering
\includegraphics[width=0.93\textwidth]{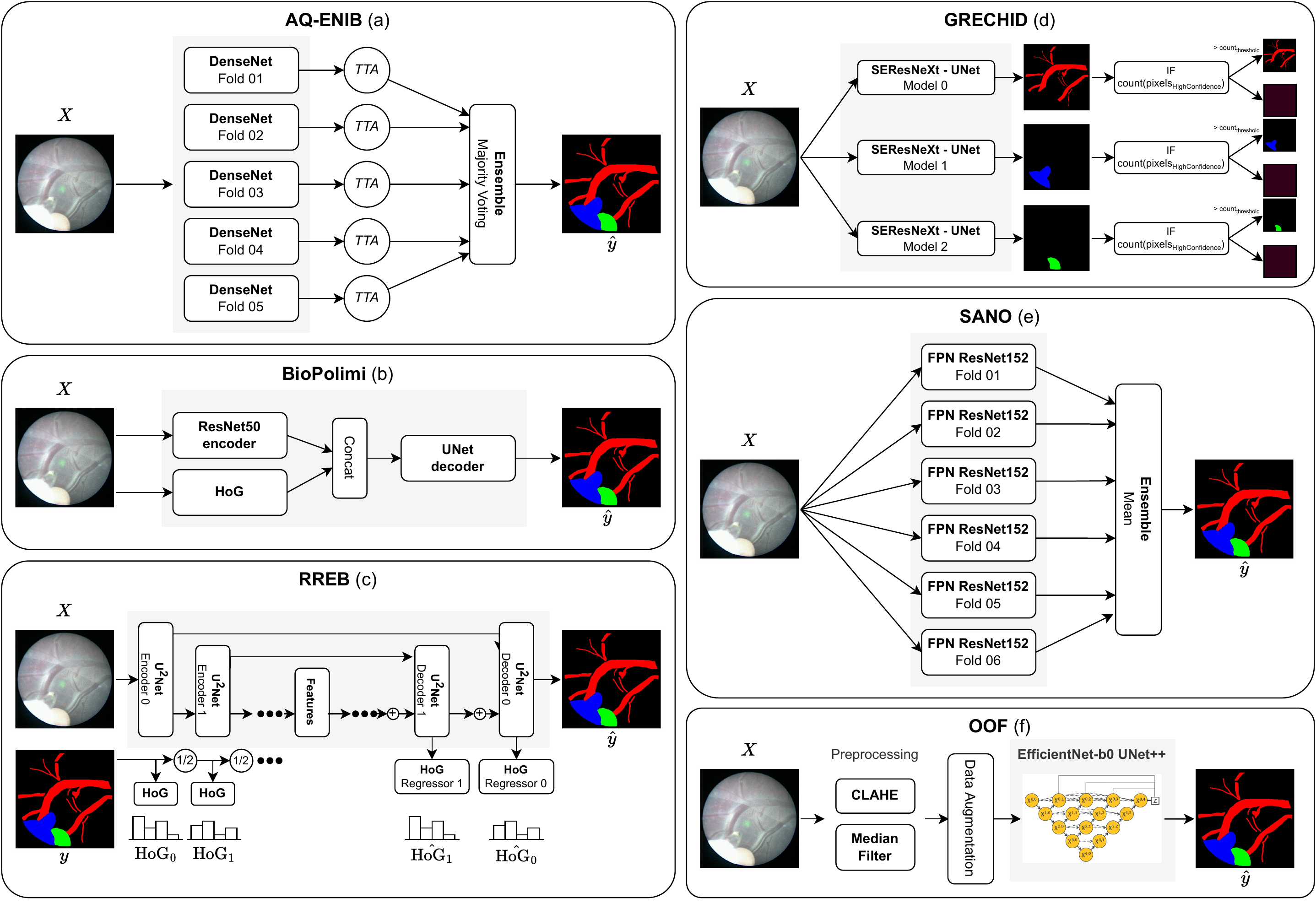}
\caption{Graphical overview of the participants' methodologies for Task 1 as described in Sec.~\ref{sec:team_summary} (Key: $X$ - input frame; $y$ - groundtruth; $\hat{y}$ - prediction). AQ-ENIB (a) proposed an ensemble of DenseNet models with Test Time Augment (TTA). BioPolimi (b) combined ResNet50 features with Histogram of oriented Gradients (HoG) computed on $X$. RREB (c) proposed a multi-task $U^{2}Net$ for segmentation and multi-scale regression of HoG features ($\hat{\textrm{HoG}_{0}}$,$\hat{\textrm{HoG}_{1}}$,...) computed on $y$ ($\textrm{HoG}_{0}$,$\textrm{HoG}_{1}$,...). GRECHID (d) used 3 SEResNeXt-UNet models individually trained on each class ensembled by thresholding, where $\textrm{pixels}_\textrm{HighConfidence}$ are pixels predicted with high confidence and $\textrm{count}_\textrm{threshold}$ is the empirical threshold. SANO (e) proposed a mean ensemble of Feature Pyramid Network (FPN) with ResNet152 backbone. OOF (f) used an EfficientNet UNet++, preprocessing images with contrast limited adaptive histogram equalization (CLAHE) and median filter.} 
\label{fig:methods_schema} 
\end{figure*}

\subsection{AQ-ENIB} 
\label{sec:team_aqenib}
Team AQ-ENIB are Abdul Qayyum, Abdesslam Benzinou, Moona Mazher and Fabrice Meriaudeau from ENIB (France), University Rovira i Virgili (Spain) and University of Bourgogne (France).
The method proposed by AQ-ENIB implemented a model made by a recursive dense encoder followed by a non-dense decoder.
Dense encoder is chosen to enable efficient features reuse, facilitating training convergence.
The dense encoder consists of 5 dense blocks, each consisting of 6 dense layers followed by a transition layer.
Each dense layer consists of 2 convolutional layers with batch normalization (BN) and ReLU activation functions. 
The first convolutional layer uses $1 \times 1$ kernels, while the second uses $3 \times 3$ kernels. 
The transition layers consist of a BN layer, a $1 \times 1$ convolutional layer, and a $2 \times 2$ average pooling layer. The transition layer helps to reduce feature-map size.
The dense blocks in the encoder have an increasing number of feature maps at each encoder stage. 
The model is trained using 5-fold cross-validation. 
To compute the final prediction, test time augmentation {(TTA)} is performed. This means that the model is fed with raw images and their augmented versions (using flipping and rotation with different angles). The model predicts, for each input, a segmentation mask. All the segmentation masks are ensembled using maximum majority voting. 

{The recursive dense architecture proposed by AQ-ENIB enables improved feature learning on the small training dataset, attenuating the chance of overfitting. Test time augmentation allows the team to increase the variability of the test set. A graphical schema of the method has been provided in Fig.~\ref{fig:methods_schema}(a)}

\subsection{BioPolimi}
\label{sec:team_biopolimi}
The team BioPolmini from Politecnico di Milano (Italy) are Chiara Lena, Ilaria Anita Cintorrino, Gaia Romana De Paolis and Jessica Biagioli. 
The model proposed by BioPolimi has a \textit{ResNet50}~\citep{he2016deep} backbone followed by the \textit{U-Net}~\citep{ronneberger2015u} decoder for segmentation. 
The model is trained for 700 epochs with 6-fold cross-validation, using learning rate and batch size of $10^{-3}$ and 32, respectively. To be consistent with the FetReg Challenge baseline, training images are resized to $448 \times 448$ pixels. 
Data augmentation, consisting of random crop with size $256 \times 256$ pixels, random rotation (in range $(-45^{\circ}, +45^{\circ})$), horizontal and vertical flip and random variation in brightness (in range $(-20\%, +20\%)$), is applied to the training data. 
%
%
During inference, testing images are cropped in patches of dimension $256 \times 256$ pixels. The final prediction is obtained by overlapping the prediction obtained for each patch with a stride equal to 8.

{BioPolimi enhances the baseline architecture by incorporating handcrafted features to address the issue of low contrast. The Histogram of Oriented Gradients (HoG) is specifically combined with features from ResNet50 to strengthen the recognition of anatomical contours, thereby supplying the decoder with a spatial prior of the features. A graphical schema of the method has been provided in Fig.~\ref{fig:methods_schema}(b).}

\subsection{GRECHID}
\label{sec:team_grechid}
Team GRECHID is Daria Grechishnikova from Moscow State University (Russia). 
The method proposed by GRECHID consists of a \textit{U-Net} model with \textit{SEResNeXt50} backbone~\citep{8578843} trained sequentially for each class (i.e., vessels, fetus and surgical tools). The \textit{SEResNeXt50} backbone contains \textit{Squeeze-and-Excitation} (SE) blocks, which allow the model to weigh adaptively each channel of SE blocks.
Before training, exact and near-duplicates were removed using an online software\footnote{https://github.com/idealo/imagededup}, obtaining 783 unique images from the original training dataset. 
Multi-label stratification split is performed to allocate images into train, test, and validation sets. All the images are resized to {$224 \times 224$} pixels. To improve model generalization, data augmentation is performed using horizontal and vertical flip, random rotation and flipping.
The model is trained using Adam optimizer and cosine annealing with restart as learning rate scheduler, with a loss that combines Dice and modified cross-entropy losses. 
The modified cross-entropy loss has additional parameters to penalize either false positives and false negatives. Training is carried out in two stages. During the first stage, the model is trained for 30 epochs with a higher learning rate of $10^{-3}$, then the learning rate is lowered to $10^{-5}$. Cosine annealing with restart scheduling is used until best convergence.

A triple threshold-based post-processing is applied on the model output to remove spurious pixels. 

{GRECHID proposes the use of a ResNeXt encoder for feature extraction. This approach aims to address the challenges of large intra-class variability and poor image quality by providing a better representation of features. Additionally, the per-class model ensemble and triple threshold post-processing help manage the high data imbalance. A graphical schema of the method has been provided in Fig.~\ref{fig:methods_schema}(d).}

\subsection{OOF - Overoverfitting}
\label{sec:team_oof}

Team OOF are Jing Jiao, Bizhe Bai and Yanyan Qiao from Fudan University (China), University of Toronto (Canada) and MicroPort Robotics. 
Team OOF used \textit{U-Net++}~\citep{zhou2018unet++} as the segmentation model. \textit{EfficientNetb-0}~\citep{tan2019efficientnet} pre-trained on the ImageNet dataset is used as \textit{U-Net++} encoder.
%
To tackle illumination variability, median blur and Contrast Limited Adaptive Histogram Equalization (CLAHE) are applied to the images before feeding them to the model.
Data augmentation, including random rotation, flip, and elastic transform, is applied during training.
Adam optimizer with an initial learning rate of $10^{-4}$ is used. The learning rate increases exponentially with 5 warm-up epochs.

{OOF addresses the issue of low contrast in images by applying the Contrast Limited Adaptive Histogram Equalization (CLAHE) technique to enhance the visibility of vessel borders. Along with visual challenges, the team encountered moiré patterns in some images that could pose difficulties in identifying the vessels. To better learn features from a small and unbalanced dataset, various configurations of EfficientNet were used as feature extractors, combined with a U-Net++ architecture and trained using standard data augmentation techniques. After evaluating the results, the team determined that the EfficientNet-b0 configuration was the best option to submit, as deeper architectures did not result in improved performance during validation. A graphical schema of the method has been provided in Fig.~\ref{fig:methods_schema}(f).}

\subsection{RREB}
\label{sec:team_rreb}

Team RREB are Binod Bhattarai, Rebati Raman Gaire, Ronast Subedi and Eduard Vazquez from University College London (UK), NepAL Applied Mathematics and Informatics Institute for Research (Nepal) and Redev Technology (UK). 
The model proposed by RREB uses \textit{$U^2$-Net}~\citep{qin2020u2} as the segmentation network. A regressor branch is added on top of each decoder layer to learn the Histogram of Oriented Gradients (HoG) at different scales. 
The loss $\mathit{L}$ minimized during the training is defined as:
\begin{equation}
    \mathit{L} =\alpha\textrm{CE}_\textrm{seg} + \beta\textrm{MSE}_{\textrm{HoG}}
\end{equation}
where $\alpha = 1$, $\textrm{CE}_\textrm{seg}$ is the cross-entropy loss for semantic segmentation, $\beta = 1$ and $\textrm{MSE}_\textrm{HoG}$ is the mean-squared error of the HoG regressor.

All the images are resized to $448 \times 448$ pixels, and random crops of $256 \times 256$ are extracted. Random rotation between $(-45^{\circ}, +45^{\circ})$, cropping at different corners and centers, and flipping are applied as data augmentation. 
The entire model is trained for 200000 iterations using Adam optimizer with $\beta_{1}=0.9$ and $\beta_{2} = 0.999$ and a batch size of $16$. 
The initial learning rate is set to $0.0002$ and then is halved at 75000, 125000, 175000 iterations. 
The proposed model is validated through cross-validation. 

{RREB team proposes the use of $U^2$-Net to enhance the learning of multi-scale features in fetoscopic images. They believe that combining handcrafted features with semantic segmentation and detection can better represent the structure of interest without incurring extra costs. To achieve this, RREB's network learns HoG descriptors as an auxiliary task, by adding regression heads to $U^2$-Net at each scale. A graphical schema of the method has been provided in Fig.~\ref{fig:methods_schema}(c).}

\begin{table*}[ht!]
\centering
\caption{{Results of segmentation on the test set for the Task 1 by training the baseline on videos only from one center. Each center ID is also indicated (I - UCLH, II - IGG) for performance comparison between the two centers.}}
\label{tab:seg_variability}
\resizebox{1.0\textwidth}{!}{
\begin{tabular}{ccccccccc}
\hline \noalign{\smallskip}
\textbf{Train Dataset} &\textbf{Video010} &\textbf{Video012} &\textbf{Video015} &\textbf{Video020} &\textbf{Video024} &\textbf{Video025} &\textbf{Overall mIoU} \\ \cline{1-7}   
\textbf{Center ID} &II &II &I &I &II &I  & \\ \hline \noalign{\smallskip}
I+II &\textbf{0.5750} &\textbf{0.4122} &\textbf{0.6923} &\textbf{0.6757} & \textbf{0.5514}	&\textbf{0.7045} &\textbf{0.6763} \\ 
I & 0.0109 & 0.0092 & 0.1012 & 0.0754 &0.0056 & 0.2180 & 0.1102 \\ 
II & 0.1968	& 0.2630 & 0.1525	&0.1562	&0.3545	&0.1907 &0.1761 \\ \hline
\noalign{\smallskip}
\end{tabular}
}
\end{table*}

\subsection{SANO}
\label{sec:team_sano}

Team SANO from Sano Center for Computational Medicine (Poland) are Szymon Płotka, Aneta Lisowska and Arkadiusz Sitek. This is the only team that participated in both tasks. 
\paragraph{Segmentation}
The model proposed by SANO is a \textit{Feature Pyramid Network} (FPN)~\citep{lin2017feature} that uses \textit{ResNet-152}~\citep{he2016deep} with pre-trained weights as backbone. The first convolutional layer has a 3-input channel, $n = 64$ feature maps, $7 \times 7$ kernel with $\mathrm{stride} = 2$, and $\mathrm{padding} = 3$. The following three convolutional blocks have $2n$,$4n$ and $32n$ feature maps. Our bottleneck consists of three convolutional blocks with BN.
During training, the images are resized to $448 \times 448$ pixels and following augmentations are applied:
\begin{itemize}
    \item color jitter (brightness = $[0.8, 1.2]$, contrast = $[0.8, 1.2]$, saturation = $[0.8, 1.2]$, and hue = $[-0.1, 0.1]$)
    \item random affine transformation (rotation = $[-90, 90]$, translation = $[0.2, 0.2]$, scale = $[1, 2]$, shear = $[-10, 10]$)
    \item horizontal and vertical flip.
\end{itemize}
The overall framework is trained with cross-entropy loss using a batch size of 4, Adam as optimizer with an initial learning rate of $10^{-4}$, weight decay and step learning rate by $0.1$, and cross-entropy loss. 
Validation is performed with 6-fold cross-validation.

{SANO propose to use a deeper feature encoder \textit{ResNet-152}, to increase the number of features extracted, on top of a FPN architecture to tackle image complexity and improve segmentation performance. A graphical schema of the strategy proposed by SANO team for Task 1 is shown in Fig.~\ref{fig:methods_schema}(e).}

\paragraph{Registration}
The algorithm uses the channel corresponding to the placental vessel (PV) from the segmentation network and the original RGB images. 
The algorithm only models translation with the precision of 1 pixel. If frames are indexed by $i = 1,\ldots,t,\ldots,T$, the algorithm finds $T - 1$ translations between neighboring frames.
To compute the placenta vasculature (PV) image, softmax is applied to the raw output of the segmentation. The PV channel is extracted and multiplied by 255. 
A mask of non-zero pixels is computed from the raw image and applied to the PV image. The homography is then computed in two steps:
The shift between PV images $t$ and $t+1$ is computed using masked Fast Fourier Transform. Then, the rotation matrix between $t$ and the shifted $t+1$ image ${T+1}_{s}$ is computed by minimizing the mean square error.

\begin{figure*}[ht!]
\centering
	\begin{subfigure}[b]{1\textwidth}
		\centering
		\includegraphics[width=0.75\textwidth]{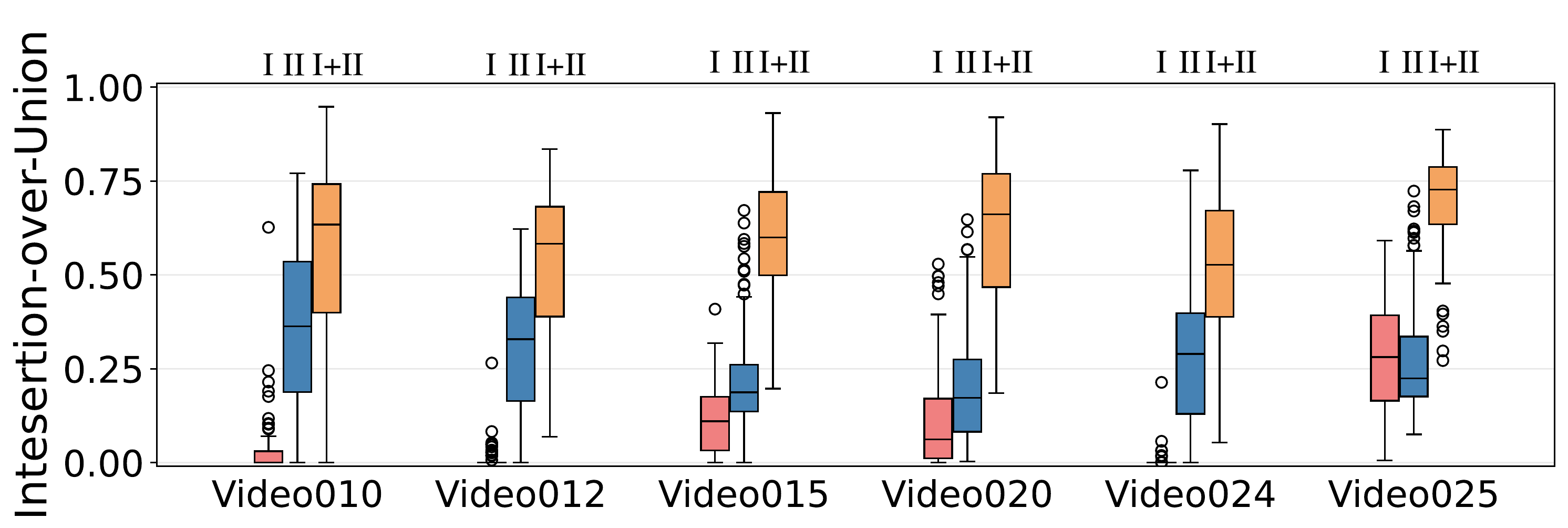}
	\end{subfigure}	
	\caption{{Qualitative comparison showing results for baseline model when trained on single center data and multi-center data. \textit{mIoU} over each test video for the baseline model trained with data from one center (I - UCLH, II - IGG). Bar colors from left to right indicate Centre I, II and I+II results.}}  
	\label{fig:segresult_variability}     
\end{figure*}

\subsection{Baseline}

As the baseline model, we trained a \textit{U-Net}~\citep{ronneberger2015u} with \textit{ResNet50}~\citep{he2016deep} backbone as described in~\cite{Bano2020deep}. 
Softmax activation is used at the final layer. Cross-entropy loss is computed and back propagated during training.  
Before training, the images are first resized to $448 \times 448$ pixels. To perform data augmentation, at each iteration step, a patch of $256 \times 256$ pixels is extracted at a random position in the image. Each of the extracted patches is augmented by applying a random rotation in range $(-45^{\circ} , +45^{\circ})$, horizontal and vertical flip, scaling with a factor in the range of $(-20\%, +20\%)$ and random variation in brightness $(-20\%, +20\%)$ and contrast $(-10\%, +10\%)$. 
Segmentation results are obtained by inference using $448 \times 446$ pixels resized input image. 
The baseline model is trained for 300 epochs on the training dataset. We create 6 folds, where each fold contains 3 procedures, to preserve as much variability as possible while keeping the number of samples in each fold approximately balanced. 
The final model is trained on the entire dataset, splitting videos in 80\% for training and 20\% for validation. The data is distributed to represent the same amount of variability in both subsets.

\section{Quantitative and Qualitative Evaluation Results}

\subsection{{Data variability contribution}}

{To assess data variability contribution from the multi-center dataset, we compute the performance of our baseline model when trained on data from one surgical center and tested on data from the other one. Table~\ref{tab:seg_variability} shows the \textit{mIoU} over each of the 6 test video samples and the overall \textit{mIoU} over all videos with the baseline model trained on dataset from a single center. Figure~\ref{fig:segresult_variability} shows (a) the qualitative comparison of mean performance over each test video for baseline model trained with data from only one center.
When training the model on data from Center I, the baseline performance on all test videos are generally lower compared to the one trained on data from Center II, except for Video025, which obtained an average mIoU of 0.1102 and 0.1761 respectively.}

{The difference in baseline model performance is mainly due to the variability and size of the dataset. In Center I, the images are higher quality and have well-visible structures. Although this is beneficial for clinicians, it needs to provide more information for the model learning process, which may lead to overfitting and poor segmentation performance. In contrast, data from Center II is more diverse, with various cases treated (e.g., different placenta positions and gestational weeks) and various imaging setups (e.g., straight or 30-degree fetoscope, brightness, FoV size). The increased image variability from these factors enables the model to generalize better to test images. Another crucial factor is that dividing the two datasets reduces the training set to about 900 images.}

{It can also be observed that when trained on individual center data, the model is not generalizable on the other center data due to data variability. 
However, combining the datasets (I+II) enhances the baseline model performance (average mIoU of 0.6763) and generalization capabilities, as it introduces a more extensive collection of images with higher variability.}

\begin{table*}[t]
\centering
\caption{{Performance of participating methods for the Task 1 (segmentation) on on the test dataset. Each center ID is also indicated (I - UCLH, II - IGG) for performance comparison between the two centers.}}
\label{tab:seg_results}
\resizebox{1.0\textwidth}{!}{
\begin{tabular}{lcccccccc}
\hline \noalign{\smallskip}
\textbf{Team name} &\textbf{Video010} &\textbf{Video012} &\textbf{Video015} &\textbf{Video020} &\textbf{Video024} &\textbf{Video025} &\textbf{Overall mIoU} &\textbf{\# Video won}  \\ \cline{1-7}   
\textbf{Center ID} &II &II &I &I &II &I  & \\ \hline \noalign{\smallskip}
AQ-ENIB &0.5611	&0.2745	&0.4855	&0.4848	&0.3342	&0.6414 &0.5503 &0 \\ 
Baseline~\citep{Bano2020deep} &\textbf{0.5750} &\textbf{0.4122} &\textbf{0.6923} &\textbf{0.6757} &0.5514	&0.7045 &\textbf{0.6763} &4 \\ 
BioPolimi &0.3891	&0.2806	&0.2718	&0.2606	&0.3666	&0.3943 &0.3443 &0 \\ 
GRECHID &0.4768	&0.3792	&0.5884	&0.5744	&0.3097	&0.6534 &0.5865 &0 \\ 
OOF &0.1874	&0.1547	&0.2745	&0.2074	&0.0872	&0.3724 &0.2526 &0 \\ 
RREB &0.5449	&0.3765	&0.6823	&0.6191	&\textbf{0.6443}	&\textbf{0.7585} &0.6411 &2 \\ 
SANO &0.4682	&0.3277	&0.5201	&0.5863	&0.4132	&0.6609 &0.5741 &0 \\ \hline \noalign{\smallskip}
\end{tabular}
}
\end{table*}

\begin{figure*}[t!]
	\begin{subfigure}[b]{0.5\textwidth}
		\centering
		\includegraphics[width=1\textwidth]{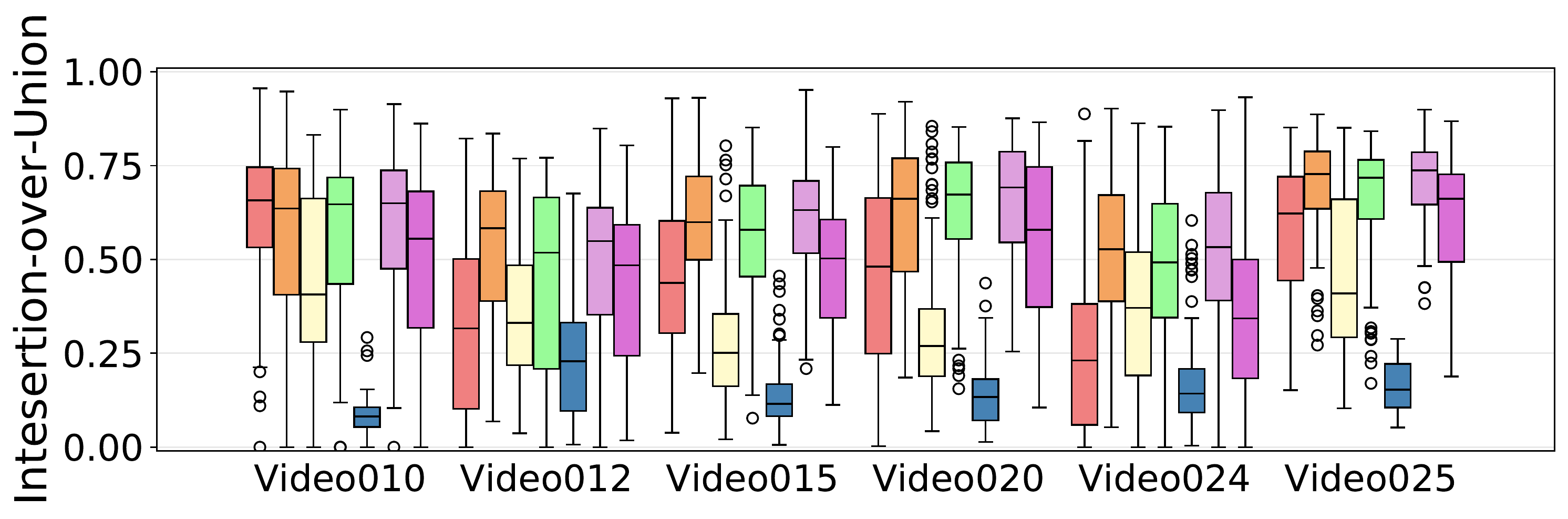}
		\caption{}	
	\end{subfigure}	
	\begin{subfigure}[b]{0.5\textwidth}
    	\centering
    	\includegraphics[width=1\textwidth]{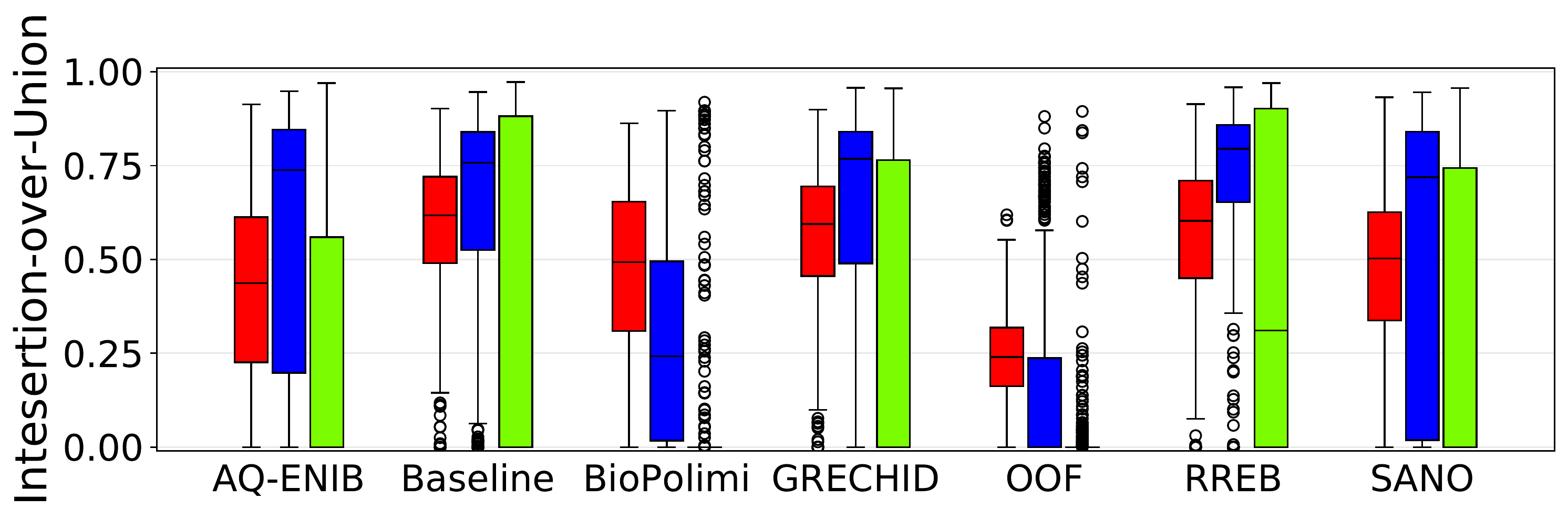}
    	\caption{}					
	\end{subfigure}	
	\caption{Qualitative comparison showing (a) \textit{mIoU} for each team on each video, and (b) overall \textit{mIoU} for each team per segmentation class. {Bar colors from left to right in (a) indicate team in alphabetical order and in (b) indicate vessel, tool, fetus classes.}}  
	\label{fig:segresults_miouperteamperclass}     
\end{figure*}

\subsection{Placental Scene Segmentation Task}
We perform both quantitative and qualitative comparison to evaluate the performance of the submitted placental scene segmentation methods. Table~\ref{tab:seg_results} shows the \textit{mIoU} for each team individually over each of the 6 test video samples and the overall \textit{mIoU} over all videos. To test the rank stability, the total number of times a team is ranked 1st on a video is also reported. Figure~\ref{fig:segresults_miouperteamperclass}(a) shows the qualitative comparison of each team on each video, and Fig.~\ref{fig:segresults_miouperteamperclass}(b) shows the comparison of each team on individual segmentation classes.

\begin{figure*}[ht]
\centering
\includegraphics[width=\linewidth]{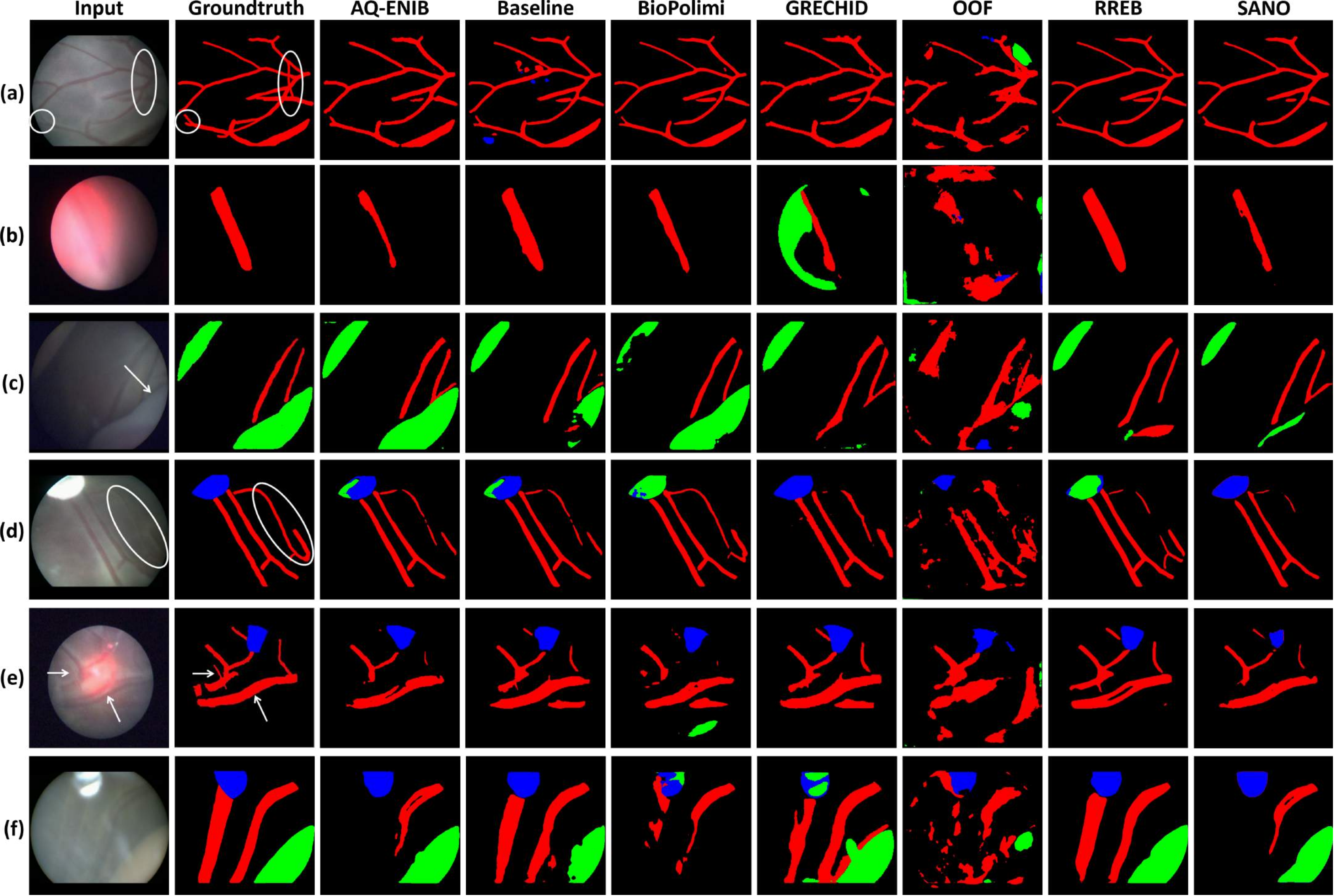}
\caption{Qualitative comparison of the 7 methods under analysis. Both baseline and RREB better generalize over the placental scene dataset. Baseline achieved better segmentation than RREB in (c), (d) and (e). OOF is the {least performing} as it failed to generalize, wrongly segmenting vessels and missed the fetus class. {White markers on the input and groundtruth images indicate regions where observations can be drawn between the seven methods under comparison.}}  
\label{fig:seg_qualitative_results}
\end{figure*}

The qualitative results for the placental scene segmentation task are presented in Fig.~\ref{fig:seg_qualitative_results}.  
Among the challenge participants, the best performing approach is that of RREB, which achieved an overall \textit{mIoU} of 0.6411. RREB obtained the best performance for all videos, but Video010 and Video012, where AQ-ENIB and GRECHID were the best, respectively.
RREB performed the best among participants for all the three classes, with median IoU for vessel and surgical tools that overcome 60\%. However, RREB obtained poor results for fetus segmentation, with a median IoU lower than 40\% with a large dispersion among images. 
As shown in Fig.~\ref{fig:seg_qualitative_results}(c) and (d), RREB meets challenges in presence of fetus and surgical tools. In the first case, RREB does not segment the fetus, while in the second the tool is segmented as fetus.

GRECHID scored second among all the participants, with a \textit{mIoU} of 0.5865. As for RREB, GRECHID grants the best and lowest performance for surgical tools and vessels, respectively.
Figure~\ref{fig:seg_qualitative_results}(b) and (f) show that GRECHID wrongly identifies and segments the fetus when it is not present in the FoV, while in Fig.~\ref{fig:seg_qualitative_results}(c), where the fetus is present, GRECHID does not segment it.

With an overall \textit{mIoU} of 0.5741, SANO scored third, with the best performance achieved for vessels. SANO shows high variability in the IoU computed among frames for both fetus and surgical tools. Despite the generalized good visual performance among videos, SANO tends to underestimate the areas. 

AQ-ENIB obtained an overall \textit{mIoU} of 0.5503 with the {least} performance obtained with fetus segmentation. Despite the good performance for vessel segmentation, vessel area is often underestimated as shown in Fig.~\ref{fig:seg_qualitative_results}(b), (e) and (f).

BioPolimi and OOF show the {least} performance with an \textit{mIoU} of 0.3443 and 0.2526, respectively. OOF also faced challenges in images where one single vessel is present in the FoV, as shown in Fig.~\ref{fig:seg_qualitative_results}({b)}. Despite the low overall performance of BioPolimi, especially in tool and fetus segmentation, vessels are correctly segmented when visible and continuous (i.e., particles or specularities does not interrupt vessels surface), as shown in Fig.~\ref{fig:seg_qualitative_results}(d).

The baseline method is the best performing method achieving an overall \textit{mIoU} of 0.6763, overcoming the performance of the challenge participants for all videos but Video024 and Video025 where RREB is the best performing method. 

\begin{table*}[t!]
\centering
\caption{Results of Registration for the Task 2 using test video clips. Mean and Median of 5-frame-SSIM metric over individual video clips is reported.}
\label{tab:reg_results}
\resizebox{1.0\textwidth}{!}{
\begin{tabular}{llcccccccc}
\hline \noalign{\smallskip}
\textbf{Team name} & &\textbf{Video010} &\textbf{Video012} &\textbf{Video015} &\textbf{Video020} &\textbf{Video024} &\textbf{Video025} &\textbf{Overall} &\textbf{\# Video won} \\ \cline{1-8}   
\textbf{Center ID} & &II &II &I &I &II &I & &   \\ \hline \noalign{\smallskip}
\multirow{2}{*}{Baseline~\citep{Bano2020deep}} &\textbf{Mean} &0.9048	&0.9204	&0.9695	&0.9169	&0.9336	&0.9558 &0.9348 &\multirow{2}{*}{5}\\ \cline{2-9} 
&\textbf{Median} &0.9303	&0.9330	&0.9767	&0.9301	&0.9478	&0.9712 &0.9524 &\\ \hline  \noalign{\smallskip} 
\multirow{2}{*}{SANO} &\textbf{Mean} &0.8231	&0.9164	&0.9588	&0.8276	&0.9420	&0.9234 &0.9019 &\multirow{2}{*}{1}\\ \cline{2-9} 
&\textbf{Median} &0.8837	&0.9289	&0.9746	&0.8825	&0.9563	&0.9608 &0.9434 &\\ \hline \noalign{\smallskip}
\end{tabular}
}
\end{table*}

\begin{figure*}[ht!]
\centering
\includegraphics[width=\linewidth]{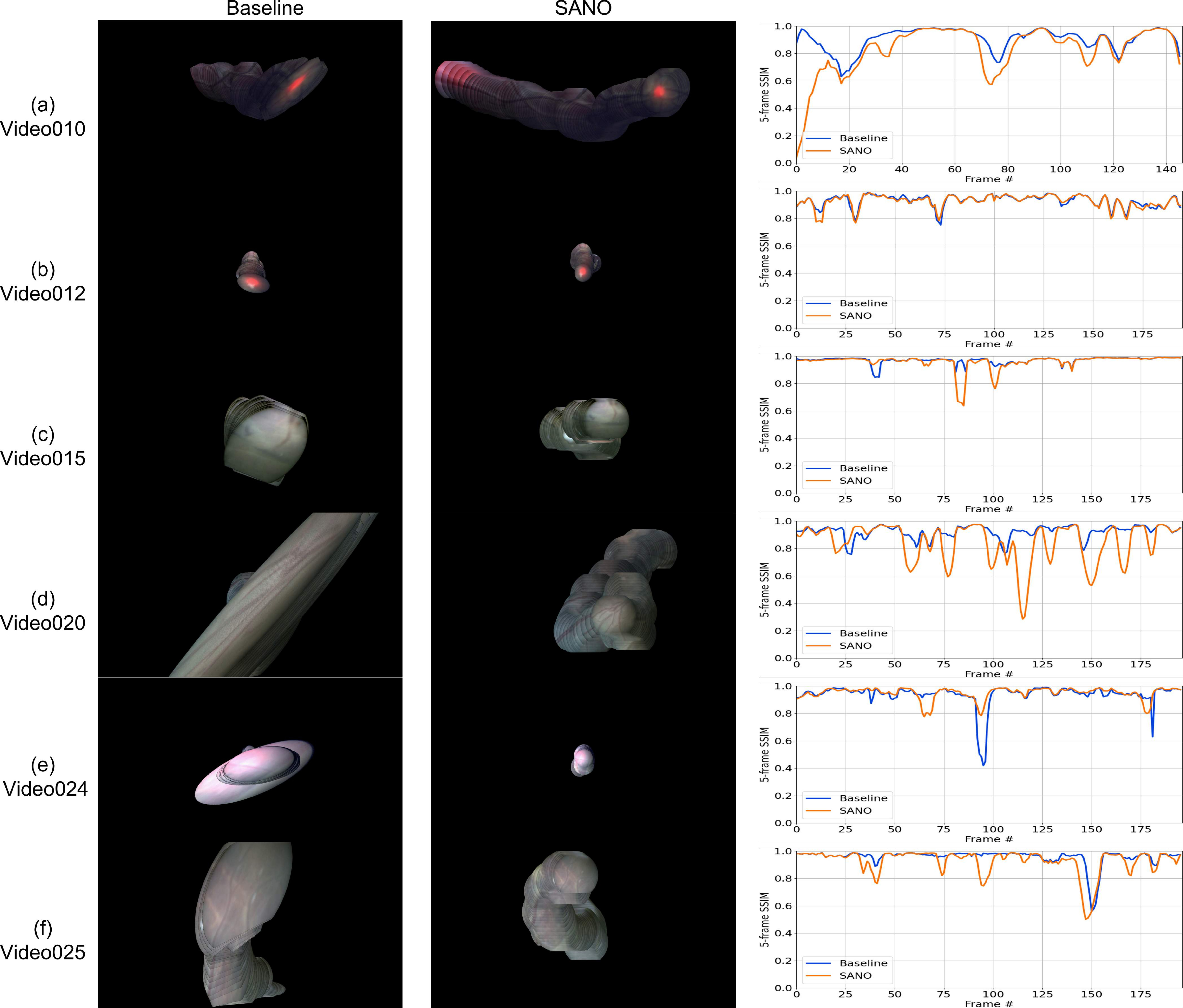}
\caption{Qualitative comparison of the Baseline~\citep{Bano2020deep} and SANO methods showing (first column) generated mosaics from the Baseline method, (2nd column) generated mosaics from the SANO method, and (3rd column) 5-frame SSIM per frame for both methods. Baseline performance is better in all videos except Video020.} 
\label{fig:reg_qualitative_results}
\end{figure*}

\begin{figure*}[ht!]
	\begin{subfigure}[b]{0.33\textwidth}
		\centering
		\includegraphics[width=1\textwidth]{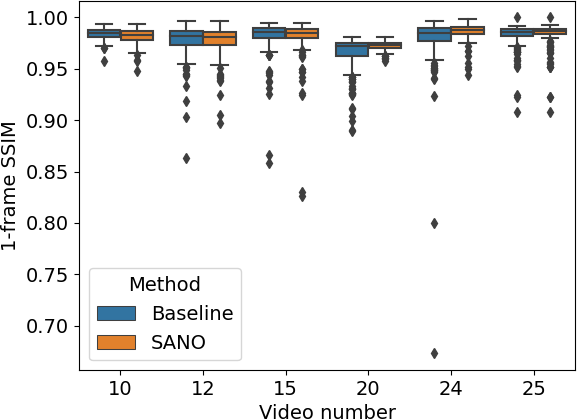}
		\caption{1-frame SSIM}	
	\end{subfigure}	
	\begin{subfigure}[b]{0.33\textwidth}
    	\centering
    	\includegraphics[width=1\textwidth]{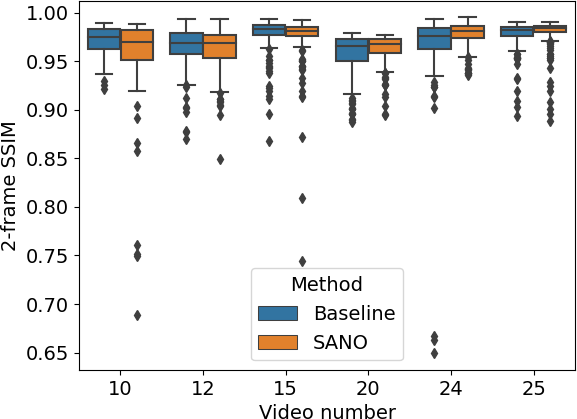}
    	\caption{2-frame SSIM}					
	\end{subfigure}	
	\begin{subfigure}[b]{0.33\textwidth}
    	\centering
    	\includegraphics[width=0.99\textwidth]{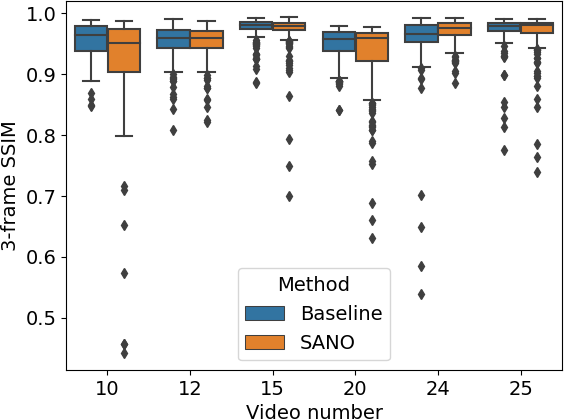}
    	\caption{3-frame SSIM}					
	\end{subfigure}	\\
	\centering
	\begin{subfigure}[b]{0.33\textwidth}
    	\centering
    	\includegraphics[width=1\textwidth]{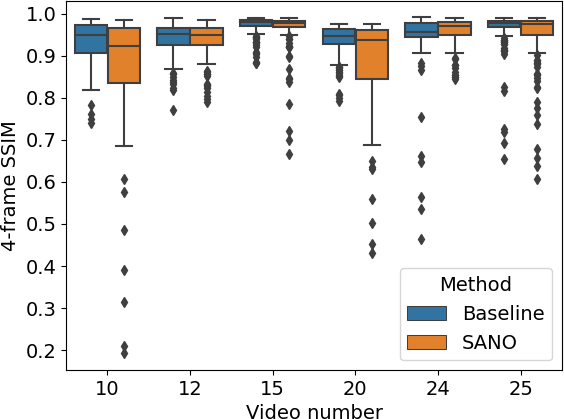}
    	\caption{4-frame SSIM}					
	\end{subfigure}	
	\begin{subfigure}[b]{0.33\textwidth}
    	\centering
    	\includegraphics[width=1\textwidth]{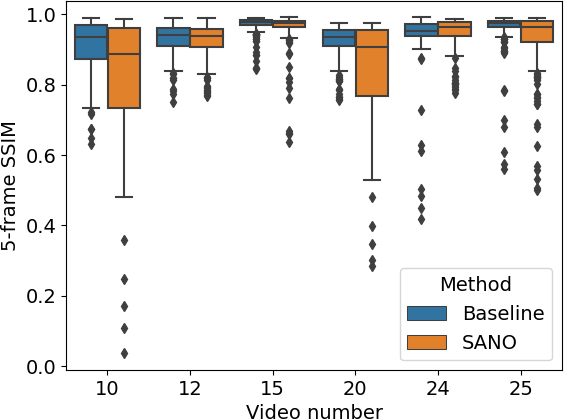}
    	\caption{5-frame SSIM}					
	\end{subfigure}	
	\caption{Quantitative comparison of the Baseline~\citep{Bano2020deep} and SANO methods using the $N$-frame SSIM metric.}  
	\label{fig:reg_quantitative_5fSSIM}     
\end{figure*}
\subsection{Registration for Mosaicking Task}
Quantitative and qualitative results for the mosaicking task are presented in Table~\ref{tab:reg_results}, Fig.~\ref{fig:reg_quantitative_5fSSIM} and Fig.~\ref{fig:reg_qualitative_results}. 

The mosaics from the baseline and SANO methods and their 5-frame SSIM metric for every pair of images 5 frames apart in a sequence are shown in Fig.~\ref{fig:reg_qualitative_results} for all 6 test video clips. Both methods utilized placental vessel maps for estimating the transformation between adjacent frames. From the mosaic of Video010, we observe that both methods followed different strategies for registration. SANO utilized translation registration having fewer degrees of freedom, while baseline performed affine registration of vessel having more degrees of freedom. Therefore, baseline is able to deal with perspective warpings while SANO's approach is unable to deal with perspective changes and overestimates translation to compensate such changes. As a result, the 5-frame SSIM for SANO is lower compared to the baseline in Video001. On Video012, both methods struggled to generate a meaningful mosaic, but overall the baseline resulted in better 5-frame SSIM metric compared to SANO (see Table~\ref{tab:reg_results}). Video015 is an anterior placenta case in which the placental surface is not fronto-parallel to the camera. As a result, there is large perspective warping across different frames. SANO's approach failed in Video015 as it estimated only translation transformation. On the other hand, the baseline successfully estimated the warping through affine transformation, resulting in better 5-frame SSIM metric. 
Qualitatively, SANO performed better on Video020 compared to the baseline, especially in regions where vessels are visible and the mosaic remained bounded due to only translation transformation estimation. However, the error between 5 frames is particularly large for SANO as the warpings are not accurate. Video024 and Video025 show interesting cases where in some frames there are no distinguishable structures like vessels (frame 90 in Video024 and frame 148 in Video025), hence both methods lost tracking intermediately. Quantitatively, SANO's performance is slightly better than the baseline on Video024. Through rank stability test, we found that baseline performance was better in 5 out of 6 videos (see Table~\ref{tab:reg_results}). 

Figure.~\ref{fig:reg_qualitative_results} shows the qualitative comparison using 1 to 5 frame SSIM metric. We observe that with increasing frame distance, the error becomes large. In the case of SANO, {Video010} and {Video015} results in large drift even at 2-frame distance. As SANO used a translation transformation estimation, its error becomes very large in all videos when observing from 1 to 5 frames SSIM. The baseline followed an affine transformation estimation, as a result, its errors appear to be relatively smaller than SANO, which mainly occurred when no visible vessels were present in the scene. 

\section{Discussion}
An accurate placental semantic segmentation is necessary for better understanding and visualization of the fetoscopic environment; as a result, this may facilitate surgeons in improved localization of the anastomoses and better surgical outcome. However, the high intra and inter-procedure variability remain a key challenge, as only a small subset of images from each procedure were manually annotated for model training. Additionally, datasets captured from different clinical centers varies in terms of the resolution, imaging device and light source, making model generalization even more challenging. From the segmentation model results on individual 6 test videos, we observed large variability in the \textit{mIoU} values of all methods (see Table~\ref{tab:seg_results}). Note that Video010, 012 and 024 are from Center II and the remaining were acquired from Center I. 
The performance of RREB, i.e., the winning team, may be explained by the use of a multi-task approach to segment anatomical structures while regressing the HoG. We hypothesize that training a CNN to regress {multi-scale} HoG {from labels enhance borders and may help the network in segmenting poorly contrasted regions}. {RREB remains the best performaning team on the tool class. Despite HoG helps in better understanding the contours and thus producing smoother segmentation masks, this does not improve the performance with non-uniform texture, as for reflections on vessel surface (Fig.~\ref{fig:seg_qualitative_results}(e) which can cause holes in the final segmentation mask, and fetus.} The poor performance of RREB for fetus segmentation may be attributed to the low number (293, i.e. 14.22\% of training frames) and high {texture} variability of fetus frames used for training.
Baseline method, which is also the top performing method, was relatively low on Center II videos (average \textit{mIoU} of 0.5130) compared to Center I videos (average \textit{mIoU} of 0.6910). Team RREB, the second-best method, also performed poorly on Video012 (mIoU of 0.3765). 

{The runner-up team, GRECHID, achieved the best performance in vessel segmentation close to RREB and baseline, with some issues in segmenting the fetus (Fig.~\ref{fig:seg_qualitative_results}(b,c,f)). GRECHID network architecture is rather similar to the baseline, but the adoption a per-class network configuration was chosen to achieve one-vs-all pixel classification and, thus eased data distribution learning for each class. While we cannot speculate whether this design actually improves the performance, it would be interesting to assess the data reduction impact on segmentation performance.}

{AQ-ENIB (average mIoU of 0.5503) and SANO (average mIoU of 0.5741) share the same segmentation strategy with only minor differences as also reflected from the comparable performance ($\Delta$mIoU of $4.32\%$). Overall, both models perform well and have the same weakness producing no or under segmentation in case of reflections (Fig.~\ref{fig:seg_qualitative_results}(c)), small vessels (Fig.~\ref{fig:seg_qualitative_results}(d)) and poor contrast (Fig.~\ref{fig:seg_qualitative_results}(f)). Test-Time Augmentation in AQ-ENIB can provide some help in fetus segmentation but also cause false positive as in~Fig.~\ref{fig:seg_qualitative_results}(d).
Considering the low difference in performance, we can analyze the models footprint and a positive aspect of AQ-ENIB is that DenseNet has lower parameters number (around 20 millions) compared to SANO ResNet152 (around 60 millions).
}

{BioPolimi uses the same architecture of the baseline but achieved way lower performance (average mIoU of 0.3443). The integration of HoG features computed on the image seemed to have a negative impact on segmentation. We hypothesize that computing HoG features on the input frame does not provide a strong reference to help network encoder to manage for low contrast, compared to HoG computed on groundtruth and multi-task as in RREB. Even though, it is worth to mention that BioPolimi method achieved the best vessel segmentation for Fig.~\ref{fig:seg_qualitative_results}(d).}

{OOF method is the least performing (average mIoU of 0.2526) on all the test set and produced several segmentation error as shown in Fig.~\ref{fig:seg_qualitative_results}. We think that the additional preprocessing generates image with high contrast, thus polarizing the network in learning non-realistic features.}

There was no single method that outperformed on all 6 test samples. This suggests that the proposed methods did not fully generalize to the dataset distributions from the two centers.
 
{Nonetheless, it is worth to consider that some of the strategies presented by the participants are complementary and can be combined to effectively tackle some of the challenges and boost the segmentation performance. Further can be performed to assess the performance by combining RREB method with other architectures, like DenseNet (AQ-ENIB) or ResNet (SANO) with GRECHID data reduction and AQ-ENIB Test-Time Augmentation. However, as stated in Sec.~\ref{sec:overview}, participants have been requested to provide their inference algorithms as Docker container thus we do not have access to their training code.}

To better model the variability in the dataset, more annotated images would be needed for supervised learning. Limited annotation problems can also be addressed through pseudo labelling using semi-supervised learning techniques.  
A reliable and consistent mosaic is needed for visualizing an increased FoV image of the placental environment. The two methods under comparison relied on accurate placental vessel segmentation for mosaicking. However, during fetoscopy, the placenta regions might appear either with very thin and weak vessels or no vessels at all. 
A segmentation algorithm may fail in these scenarios, especially when no vessels are visible, leading to failure in consecutive frames registration for mosaicking. This suggests that a registration algorithm should not solely rely on vessel segmentation predictions. More recent deep learning-based keypoint and matching approaches~\citep{detone2018superpoint, sarlin2020superglue, sun2021loftr} could be useful in improving placental frame registration for mosaicking. {Some recent works~\citep{casella2022learning,bano2022placental} in mosaicking have already shifted interests towards exploiting learning-based keypoints and matching approaches.  }

\section{Conclusion}
SDS has the potential to enhance intraoperative imaging by providing better visualization of the surgical environment with increased FoV to support the surgeon's decision during the procedure. Deep learning-based semantic segmentation algorithms can help in better understanding the fetoscopic placental scene during fetoscopy. However, large labelled datasets are required for training robust segmentation models. 
Through the FetReg2021 challenge, which was part of the MICCAI2021 Endoscopic vision challenge, we contributed a large scale multi-center fetoscopy dataset containing data from 18 fetoscopy procedures for training and 6 fetoscopy procedures for testing. The test data was hidden from the challenge participants but followed similar distribution to the training dataset. 
The challenge focused on solving the task of placental semantic segmentation and fetoscopy video frame registration for mosaicking. The segmentation solutions presented by the participating teams achieved promising results though they were unable to beat the baseline method. Achieving generalizability remained an open question, and none of the methods outperformed in all test video samples. 
The contributed mosaicking approaches relied on accurate vessel segmentation and the presence of vessels in the fetoscopic placental view. Through the FetReg2021 challenge, we contributed a benchmark dataset for advancing the research in fetoscopic mosaicking. 

\section*{Acknowledgment}
We are grateful to NVIDIA, Medtronic and E4 Computing for sponsoring the FetReg2021 challenge. 
This work was supported by the Wellcome/EPSRC Centre for Interventional and Surgical Sciences (WEISS) at UCL (203145Z/16/Z), the Engineering and Physical Sciences Research Council (EP/P027938/1, EP/R004080/1,EP/P012841/1, NS/A000027/1), the Royal Academy of Engineering Chair in Emerging Technologies Scheme, Horizon 2020 FET Open (863146) and Wellcome [WT101957]. Anna L. David is supported by the NIHR UCLH Biomedical Research Center. For the purpose of open access, the author has applied a CC BY public copyright license to any author accepted manuscript version arising from this submission.

\bibliographystyle{model2-names.bst}\biboptions{authoryear}
\bibliography{refs}



\end{document}